\documentclass[12pt,a4paper]{article}
\pdfoutput=1

\usepackage{cite,mathtools,amsmath,amssymb,comment}
\usepackage{graphicx}
\usepackage[usenames]{color}
\usepackage[bookmarksopen,colorlinks=true,linkcolor=dark_green,citecolor=dark_red,urlcolor=dark_red,linktocpage=false]{hyperref}

\definecolor{dark_blue}{rgb}{0,0,0.6}
\definecolor{dark_green}{rgb}{0,0.4,0}
\definecolor{dark_red}{rgb}{0.6,0,0}

\usepackage[height=22.5cm,width=16.5cm,centering]{geometry}

\def\thefootnote{\fnsymbol{footnote}}

\renewcommand{\thefootnote}{\fnsymbol{footnote}}
\begin{document}

\begin{titlepage}

\begin{center}

\hfill DESY 19-067 \\
\hfill CTPU-PTC-19-12 \\
\hfill KEK-TH-2125 \\

\vskip 1.5cm

{\fontsize{15pt}{0pt} \bf
Gravitational waves from first-order phase transitions:
}
\vskip 0.5cm
{\fontsize{15pt}{0pt} \bf
Ultra-supercooled transitions and the fate of relativistic shocks
}

\vskip 1.2cm

{\large
Ryusuke Jinno$^{a,b}$, Hyeonseok Seong$^{c,b}$, 
\\ \vspace{0.3cm} 
Masahiro Takimoto$^{d,e}$ and Choong Min Um$^{c}$
}

\vskip 0.5cm

\begin{tabular}{ll}
$^{a}$ &\!\! 
{\em Deutsches Elektronen-Synchrotron DESY, 22607 Hamburg, Germany} \\[.3em]
$^{b}$ &\!\! {\em Center for Theoretical Physics of the Universe, Institute for Basic Science (IBS),} \\
&{\em Daejeon 34126, Korea} \\[.3em]
$^{c}$ &\!\! {\em Department of Physics, KAIST, Daejeon 34141, Korea} \\[.3em]
$^{d}$ &\!\! {\em Department of Particle Physics and Astrophysics, Weizmann Institute of Science,} \\
&{\em Rehovot 7610001, Israel} \\[.3em]
$^{e}$ &\!\! {\em Theory Center, High Energy Accelerator Research Organization (KEK),} \\
&{\em Oho, Tsukuba, Ibaraki 305-0801, Japan} \\[.3em]
\end{tabular}

\end{center}
\vskip 1cm

\begin{abstract}
We study the gravitational wave (GW) production in extremely strong first order phase transitions
where the latent heat density dominates the plasma energy density, $\alpha \gtrsim 1$.
In such transitions, bubbles develop extremely thin and relativistic fluid configurations,
resulting in strong shock waves after collisions.
We first propose a strategy to understand the GW production in such a system by separating the problem into 
the propagation part and the collision part.
Focusing on the former, we next develop an effective theory for the propagation of the relativistic fluid shells.
Using this effective theory, we finally calculate the expected duration of the relativistic fluid configurations
and discuss its implications to the GW production.
\end{abstract}

\end{titlepage}

\tableofcontents
\thispagestyle{empty}

\renewcommand{\thepage}{\arabic{page}}
\setcounter{page}{1}
\renewcommand{\thefootnote}{$\diamondsuit$\arabic{footnote}}
\setcounter{footnote}{0}

\newpage
\setcounter{page}{1}

\section{Introduction}
\label{sec:Intro}
\setcounter{equation}{0}

In the upcoming decades, gravitational waves will prove to be an invaluable tool in probing the early universe.
They will provide us opportunities to test high-energy physics theories such as
inflation~\cite{Starobinsky:1979ty}, 
cosmological first-order phase transitions~\cite{Witten:1984rs,Hogan:1986qda},
preheating~\cite{Khlebnikov:1997di}, topological defects~\cite{Vilenkin:2000jqa}, and so on.
To observe gravitational waves,
the LIGO Scientific Collaboration~\cite{TheLIGOScientific:2014jea} and Virgo Collaboration~\cite{TheVirgo:2014hva} 
are now in operation, and KAGRA~\cite{Somiya:2011np} is also expected to join soon.
In addition, space-based detectors such as LISA~\cite{Seoane:2013qna} (and its possible successor BBO~\cite{Harry:2006fi}) 
and DECIGO~\cite{Seto:2001qf} have been proposed.

Among various sources of gravitational waves, cosmological first-order phase transitions are of our interest in this paper.
Although both the electroweak and QCD phase transitions are known to be crossovers
in the Standard Model (SM)~\cite{Kajantie:1996mn,Gurtler:1997hr,Csikor:1998eu}, 
first-order phase transitions often occur in a wide range of extensions of the SM,
and may play a crucial role in explaining the baryon asymmetry 
of the universe~\cite{Kuzmin:1985mm,Cohen:1993nk,Riotto:1999yt,Morrissey:2012db}.
During a first-order phase transition, bubbles of the broken phase nucleate, expand, 
and eventually collide with each other (Fig.~\ref{fig:Fluid}).\footnote{
See e.g. Refs.~\cite{Kosowsky:1991ua,Kosowsky:1992rz,Kosowsky:1992vn,Kamionkowski:1993fg} 
for pioneering works on the GW production in such systems.
}
The properties of the transitions are characterised by several quantities, 
one of which is the ratio between the latent heat and the plasma energy density just before the transition:
\begin{align}
\alpha	
&\equiv
\frac{\rm (vacuum~energy~density~released~in~the~transition)}{\rm (plasma~energy~density~in~the~symmetric~phase)}.
\end{align}
For small $\alpha$ ($\ll 1$), the dynamics of the bubbles and the resulting GW production are relatively well-known
(though still far from complete):
the released latent heat turns into heat and the kinetic energy of the surrounding plasma, 
which in turn propagates in the form of sound waves characterised by 
the linearity of the fluid equation $(\partial_t^2 - c_s^2 \nabla^2) \vec{v} = 0$.
These sound waves propagate even after bubble collisions and damping of the scalar configurations,
and can be a long-lasting source of gravitational waves~\cite{Hindmarsh:2013xza,Hindmarsh:2015qta,Hindmarsh:2017gnf,Hindmarsh:2016lnk}.
At late times, these sound waves turn into turbulence which produce another characteristic form of 
the GW spectrum~\cite{Kamionkowski:1993fg,Kosowsky:2001xp,Nicolis:2003tg,Caprini:2006jb,Caprini:2009yp,Kahniashvili:2009mf} (see also reviews e.g. Refs.~\cite{Caprini:2015zlo,Cai:2017cbj,Mazumdar:2018dfl}).

In contrast, the dynamics of the transitions with large $\alpha$ ($\gtrsim 1$) -- what we call the ultra-supercooled transitions in this paper -- is rather unknown,
despite that they are quite interesting from an observational viewpoint.\footnote{
See Refs.~\cite{Ellis:2018mja,Ellis:2019oqb} for a recent discussion on the maximal possible strength of the transitions 
in specific models.
}
Indeed, first-order phase transitions in a certain class of models are known to have 
this property~\cite{Randall:2006py,Espinosa:2008kw,Konstandin:2011dr,Hambye:2013sna,Jaeckel:2016jlh,Jinno:2016knw,Marzola:2017jzl,Iso:2017uuu,Chiang:2017zbz,vonHarling:2017yew,Bruggisser:2018mus,Bruggisser:2018mrt,Baldes:2018emh,Hashino:2018wee,Prokopec:2018tnq,Brdar:2018num,Marzo:2018nov,Baratella:2018pxi,Fairbairn:2019xog}
and we, at least naively, expect a large amount of gravitational wave to be produced because of the huge latent heat released in the universe.
Previously, it had been thought that the scalar walls start to run away with sufficiently large $\alpha$~\cite{Bodeker:2009qy},
but recently it has been pointed out that particle-splitting processes prohibit runaway solutions 
at least when the scalar field is gauged~\cite{Bodeker:2017cim}.\footnote{
Nevertheless, the scalar field can play an important role in the GW production in extremely strong transitions.
In such cases, the resulting GW spectrum has been estimated by the so-called 
envelope approximation~\cite{Kosowsky:1991ua,Huber:2008hg,Jinno:2016vai,Jinno:2017fby},
but the effects beyond the envelope approximation can also be 
important~\cite{Weir:2016tov,Jinno:2017ixd,Konstandin:2017sat,Cutting:2018tjt}.
}
This means that an extremely strong detonation is realized in such a large $\alpha$ transition,
and therefore the initial fluid profile after bubble collision is quite different from that of
sound waves (characterised by linearity $(\partial_t^2 - c_s^2 \nabla^2) \vec{v} = 0$) or turbulence.
Even if the system were to develop into sound waves or turbulence at late times,
we must first understand its evolution just after bubble collision
to know the resulting GW spectrum from this type of transition:
\begin{align}
\boxed{
\begin{matrix}
{\rm ~Extremely~thin~and~relativistic~}
\\[1.5ex]
{\rm ~fluid~profiles~}
\end{matrix}
}
~~~~~
&\to
~~~~~
\begin{matrix}
\scalebox{2}{\rm ?}
\end{matrix}
~~~~~
\to
~~~~~
\boxed{
\begin{matrix}
{\rm ~Sound~waves~(?)}
\\[1.5ex]
{\rm ~Turbulence~(?)}
\end{matrix}
}
\end{align}
Indeed, the spectrum of the gravitational waves generated from thin and long-lived bubbles 
can be quite different from that from the overlapping sound shells~\cite{Hindmarsh:2016lnk,Jinno:2017fby}.
Therefore, in order to correctly predict the GW spectrum from ultra-supercooled transitions,
we need to understand the evolution of thin and relativistic fluid profiles.
One may consider numerical simulations, but difficulties arise from
(see also the profiles in Fig.~\ref{fig:Fluid} or \ref{fig:Propagation}):
\begin{itemize}
\item
Strong energy and momentum concentration in the fluid and the resulting huge hierarchy between the thickness of the profile and 
the size of the whole system (typically $\gtrsim {\mathcal O}(10)$ $\times$ (bubble size at collisions))
\item
Strong shock waves forming around the propagating fluid
\end{itemize}
Given this, in this paper we tackle the ultra-supercooling regime and the resulting GW production with an analytic approach,
in line with the present authors' investigation so far~\cite{Jinno:2016vai,Jinno:2017fby,Jinno:2017ixd}.

The paper is organized as follows.
In Sec.~\ref{sec:Review}, we review the hydrodynamics around the expanding bubble walls before collisions.
In Sec.~\ref{sec:Strategy}, we make clear the problems we tackle
and give an overview of the strategy to deal with them.
In Sec.~\ref{sec:Effective}, we develop an effective theory for the relativistic shock propagation and check its validity by comparing it with the result of numerical simulations. 
In Sec.~\ref{sec:Implications}, we discuss the implications of our results to the GW production in large $\alpha$ transitions.
In Sec.~\ref{sec:DC}, we summarize.
In Appendix~\ref{app:Beyond}, we derive the equations without the relativistic limit,
while in Appendix~\ref{app:Detail} we explain the details of numerical simulations.

\section{Brief review on hydrodynamics around the bubble}
\label{sec:Review}
\setcounter{equation}{0}

In this section, we briefly review the fluid dynamics around an expanding bubble.
This section closely follows Ref.~\cite{Espinosa:2010hh}.
See this reference for full derivations and more detailed explanations.

\subsection{Basic equations}
\label{subsec:Basic}

\subsubsection{Energy-momentum conservation}
\label{subsec:EM}

All the necessary equations for the fluid profile before collision 
are obtained from the energy-momentum conservation
\begin{align}
\partial_\mu T^{\mu \nu}
&= 0.
\label{eq:EM}
\end{align}
The energy-momentum tensor $T_{\mu \nu}$ has two contributions,
one from the scalar field $\phi$ driving the transition $T^\phi_{\mu \nu}$
and the other from the plasma particles $T^{\rm plasma}_{\mu \nu}$.
Since we are interested in the dynamics of the cosmological-size bubbles,
the plasma particles can be well-approximated by the fluid description.
Assuming the perfect fluid form, we may write the corresponding part as
\begin{align}
T^{\rm plasma}_{\mu \nu}
&= 
w u_\mu u_\nu - p g_{\mu \nu},
\end{align}
where $w = \rho + p$ is the enthalpy density with $\rho$ and $p$ the energy density and pressure, respectively, $u^\mu = \gamma (1, \vec{v})$ is the fluid four-velocity with $\gamma = 1/\sqrt{1 - |\vec{v}|^2}$ the gamma factor of the fluid, and $g_{\mu \nu} = {\rm diag}(+,-,-,-)$ is the metric tensor.

\subsubsection{Boundary condition at the scalar wall}
\label{subsec:Scalar}

The bubbles that produce gravitational waves are of cosmological scale.
In contrast, the scalar field dynamics is restricted within the ``particle-physics scales" 
(say, $\sim$ TeV${}^{-1}$ size for a TeV scale transition), 
a configuration that can be regarded as an infinitesimal boundary separating the symmetric phase $+$ and the broken phase $-$.
The energy-momentum conservation across the boundary gives the following two conditions:
\begin{align}
w_+ v_+^2 \gamma_+^2 + p_+
&=
w_- v_-^2 \gamma_-^2 + p_-,
~~~~~~
w_+ v_+ \gamma_+^2
=
w_- v_- \gamma_-^2,
\label{eq:EMconservation}
\end{align}
where $v_\pm$ are the incoming and outgoing fluid velocities measured from the wall, respectively.

Using the relativistic gas approximation for simplicity, the pressure and energy density can be written in terms of temperature $T$ 
and vacuum energy $\epsilon$ of the symmetric phase relative to that of the broken phase:
\begin{align}
p_+
=
\frac{1}{3} a_+ T_+^4 - \epsilon,
&~~~~~~
\rho_+
=
a_+ T_+^4 + \epsilon,
\label{eq:bag_+}
\\
p_-
=
\frac{1}{3} a_- T_-^4,
&~~~~~~
\rho_-
=
a_- T_-^4,
\label{eq:bag_-}
\end{align}
where $a_\pm$ are the number of light degrees of freedom in the symmetric and broken phase, respectively.
Eqs.~(\ref{eq:EMconservation})--(\ref{eq:bag_-}) give us two relations between $v_\pm$
and two dimensionless quantities from three dimensionful quantities, $T_\pm$ and $\epsilon$:
\begin{align}
v_+ v_-
&=
\frac{1 - (1 - 3 \alpha_+)r}{3 - 3(1 + \alpha_+)r},
~~~~~~
\frac{v_+}{v_-}
=
\frac{3 + (1 - 3 \alpha_+)r}{1 + 3(1 + \alpha_+)r},
\label{eq:vpm}
\end{align}
where
\begin{align}
\alpha_+
&\equiv
\frac{\epsilon}{a_+ T_+^4},
~~~~~~
r
\equiv
\frac{a_+ T_+^4}{a_- T_-^4}.
\end{align}
Note that the first equation gives $v_+ v_- = 1/3$ for $\alpha_+ = 0$.
This will be important when we discuss the position of the shock front later.

\subsubsection{Fluid equations}
\label{subsec:Fluid}

\begin{figure}
\begin{center}
\includegraphics[width=0.49\columnwidth]{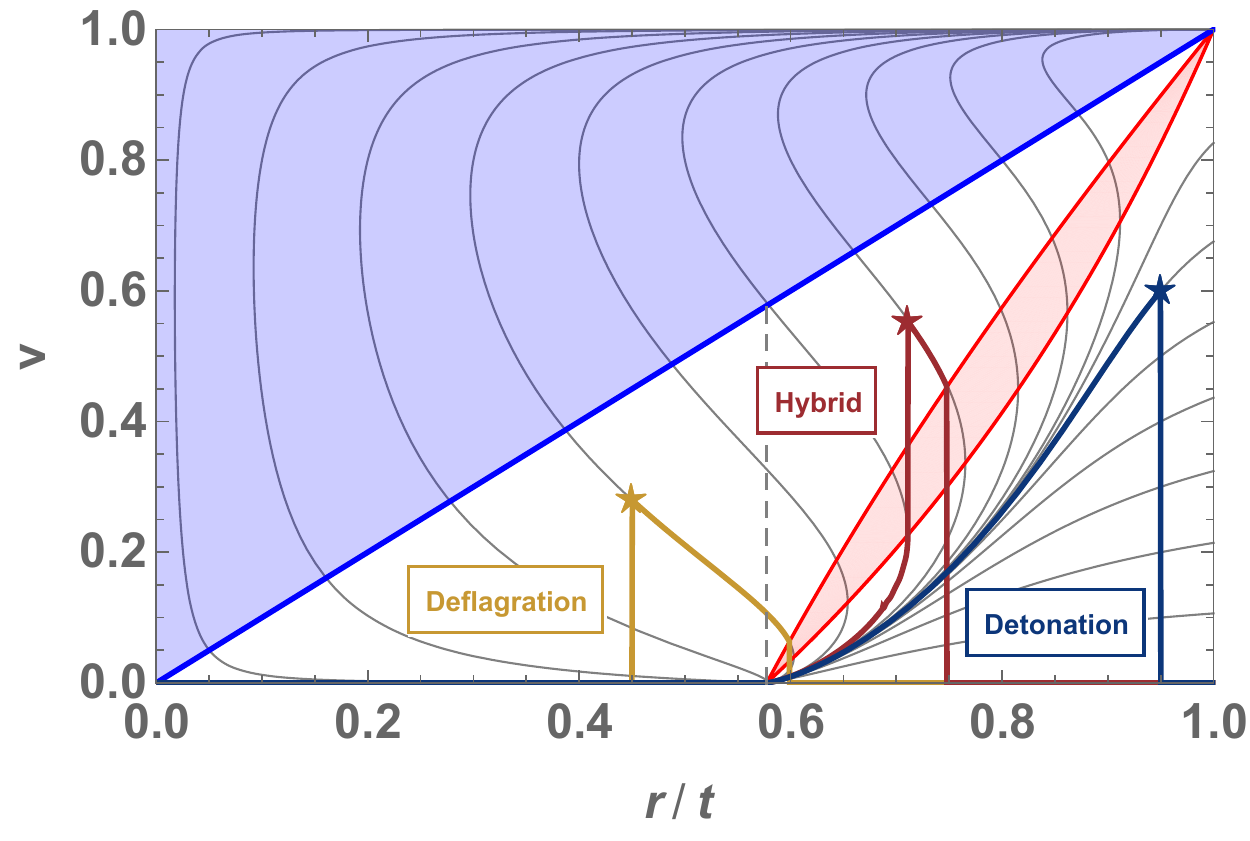}
\includegraphics[width=0.49\columnwidth]{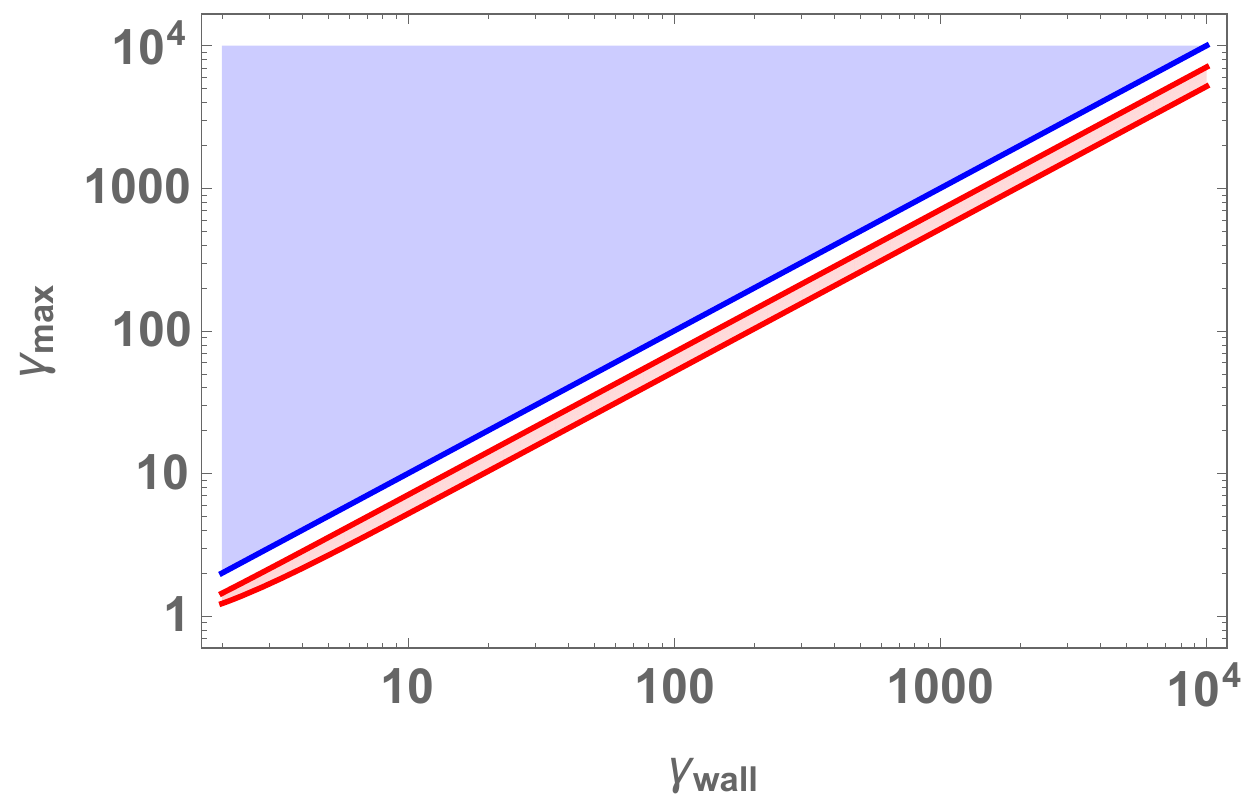}
\caption{\small
Examples of the fluid profile during bubble expansion (left)
and a closer look at the left panel's top-right corner in terms of relativistic $\gamma$ factors (right).
Stars denote both the wall positions and fluid maximal velocities.
Walls are forbidden to be inside the blue or red regions.
}
\label{fig:Profile}
\end{center}
\end{figure}

Except at the wall position, the fluid dynamics is governed by the energy-momentum conservation involving the fluid only.
Since there is no characteristic length scale in the present system,
the conservation equation $\partial^\mu T^{\rm fluid}_{\mu \nu} = 0$ can be written in terms of the velocity of a coordinate point in the wave profile, $\xi \equiv r / t$,
where $r$ is the radial coordinate and $t$ is the time elapsed since the nucleation of the bubble.
The point at the wall position reaches the terminal velocity $\xi = \xi_{\rm wall}$. $v(\xi)$ is then the fluid velocity at $\xi$ as measured from the bubble center.
After projecting onto the directions along and perpendicular to the flow of the fluid, we obtain the following equations
\begin{align}
(\xi - v) \frac{\partial_\xi \rho}{w} 
&=
2\frac{v}{\xi} + \gamma^2 (1 - v \xi) \partial_\xi v,
\\
(1 - v \xi) \frac{\partial_\xi p}{w} 
&=
\gamma^2 (\xi - v) \partial_\xi v.
\end{align}
The two derivatives $\partial_\xi \rho$ and $\partial_\xi p$ are related through the speed of sound in the plasma
$c_s^2 = (dp/dT)/(d\rho/dT)$\footnote{In this paper, we use $c_s^2 = 1/3$ in both the symmetric and broken phases assuming that all the qualitative features of the system do not depend significantly on the possible deviations from this value.}, and hence we get the following equation describing the velocity profile:
\begin{align}
\partial_\xi v
&=
\frac{2v}{\displaystyle \gamma^2 (1 - v \xi) \xi \left( \frac{\mu^2}{c_s^2} - 1 \right)}
\label{eq:v_xi}
\end{align}
where $\mu$ is the Lorentz-transformed fluid velocity measured from $\xi$:
\begin{align}
\mu (\xi, v)
&\equiv
\frac{\xi - v}{1 - \xi v}.
\end{align}
Solving Eq.~(\ref{eq:v_xi}) for $\xi$ as a function of $v$ gives the trajectories (gray lines) shown in the left panel of Fig.~\ref{fig:Profile}.
The boundary conditions at the wall positions are denoted by stars.
The $v = \xi$ line separates the blue and white regions,
while the red region is enclosed by
$\mu \xi = c_s^2$ (the upper boundary) and $\mu = c_s$ (the lower boundary).
Note that the derivative $\partial_\xi v$ diverges at the lower boundary along the gray lines. 
As seen from this figure, this system allows several types of solutions which we turn to next.

\subsection{Detonation, deflagration and hybrid}
\label{subsec:DDH}

There are three types of solutions for the motion of plasma in the present system: detonation, deflagration, and hybrid.

\subsubsection{Detonation}
\label{subsec:Detonation}

First, let us take the boundary condition (i.e. the wall position) to be somewhere below the red region in Fig.~\ref{fig:Profile}. 
An example profile is outlined in blue in Fig.~\ref{fig:Profile}. 
In this case, the differential equation (\ref{eq:v_xi}) can be easily solved to give solutions of detonation type.
In detonations, 
the supersonic wall precedes the fluid in motion, and a nonzero fluid velocity arises from $c_s < \xi < \xi_{\rm wall}$
where $\xi_{\rm wall}$ is the wall velocity as mentioned previously.

\subsubsection{Deflagration}
\label{subsec:Deflagration}

Next, let us take the boundary condition in the white region satisfying $\xi < c_s$. An example is shown in yellow in Fig.~\ref{fig:Profile}, which has the fluid moving in front of the wall. These solutions are of deflagration type.

The deflagration profiles have a shock front.
To see this, first note that all the gray lines corresponding to deflagrations cross the $\mu = c_s$ line 
(the lower boundary between the red and white regions) where the derivative $\partial_\xi v$ diverges.
As we follow the lines further down to the attractor point $(\xi,v) = (c_s,0)$, the profiles become double-valued.
Since this is unphysical, we need to devise a method to have the fluid velocities jump to zero.
The solution is to consider a shock front developing in the front end of such profile.
The location of the shock is given by a consideration similar to Eq.~(\ref{eq:vpm}):
we can redefine the quantities with the subscripts $\pm$ to describe the fluid in front of and behind the shock front, respectively,
with $\alpha_+$ taken to be $0$ since there is no energy injection at the shock position.
This gives us $v_+ v_- = 1/3$, which translates into $\mu \xi = 1/3$ 
(the upper boundary between the red and white regions) in the plasma frame.
Therefore, the fluid velocity is nonzero behind the shock front, i.e. the crossing point of the gray line and the boundary $\mu \xi = 1/3$,
while it drops to zero in front of the shock front.

\subsubsection{Hybrid}
\label{subsec:Hybrid}

Finally, let us take the boundary condition in the white region above the red region satisfying $\xi > c_s$.
An example is shown in red in Fig.~\ref{fig:Profile}.
These solutions are called the hybrid type because a nonzero fluid motion arises both in front of and behind the wall.

The fluid profile in front of the wall is similar to that of deflagrations:
the fluid velocity takes its maximum value at the wall position and vanishes at the shock front position, $\mu \xi = 1/3$.
On the other hand, the profile behind the wall is clearly different from that of deflagrations:
a nonzero fluid motion also arises behind the wall, a rarefaction wave of Jouguet type, starting from the line $\mu = c_s$.
In fact, the single-valuedness of the fluid profile requires that the profile to be on or below the $\mu = c_s$ line, but it must also have its endpoint on or above the $\mu = c_s$ line lest the fluid velocity measured from the wall exceed the speed of sound, presenting another discontinuity.

\section{Problem and basic strategy}
\label{sec:Strategy}
\setcounter{equation}{0}

\begin{figure}
\begin{center}
\includegraphics[width=0.8\columnwidth]{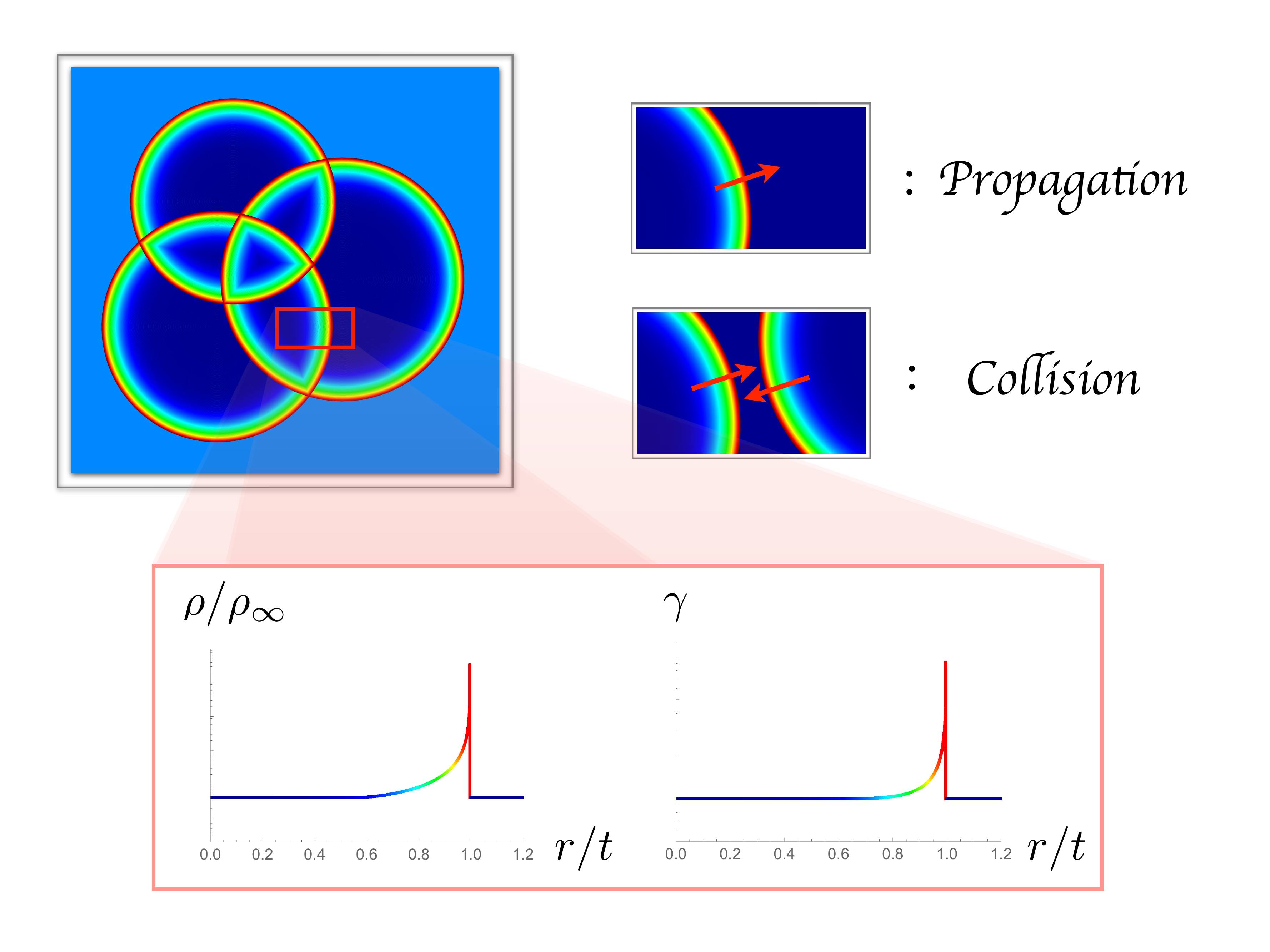} 
\caption{\small
A rough sketch of the system we consider in this paper.
In large $\alpha$ transitions, the relativistic fluid is intensely localized around the wall and has highly relativistic velocities.
After bubble collision, the fluid is pushed into the broken phase of another bubble.
}
\label{fig:Fluid}
\end{center}
\vskip 1cm
\end{figure}

Now let us focus on the fluid dynamics and the GW production in large $\alpha$ transitions. 
A sketch of the system is shown in Fig.~\ref{fig:Fluid}. 
Just after bubble collisions, the relativistic fluid is pushed inside the regions of broken phase.
To understand the evolution of such fluid, we first propose reducing the problem to two parts:
\begin{itemize}
\item
Fluid propagation
\item
Fluid collision
\end{itemize}
each of which deforms the initial profile.
We focus on the former in this paper and will report on the latter elsewhere.
Note that even the first part is highly nontrivial because of the nonlinearity of the fluid equations of motion
and the strong shock waves developing in the front end of the propagating fluid.

Let us take a closer look at the initial fluid profiles.
As explained in Sec.~\ref{sec:Review}, 
the left panel of Fig.~\ref{fig:Profile} shows the fluid velocity around an expanding bubble.
The horizontal axis is the distance measured from the bubble center normalized by the time elapsed from nucleation.
The wall positions, denoted by stars, determine the location of the energy release and the strength of the fluid dragging.
As explained previously, the profiles can be classified into three types depending on the wall position.
From a particle physics point of view, the wall position is determined by the balance between the
(1) pressure on the wall exerted by the energy release,
and 
(2) friction on the wall arising from the coupling to the plasma particles.

The left panel shows that the walls with larger energy release $\alpha$ are located closer to the top-right corner.
Since we are concentrating on the large $\alpha$ cases, the relevant combustion modes are detonations and hybrids.
In the following we mainly work with detonations to make the analysis simple. 
We discuss hybrids in Secs.~\ref{sec:Implications} and \ref{sec:DC}.

\begin{figure}
\begin{center}
\includegraphics[width=0.85\columnwidth]{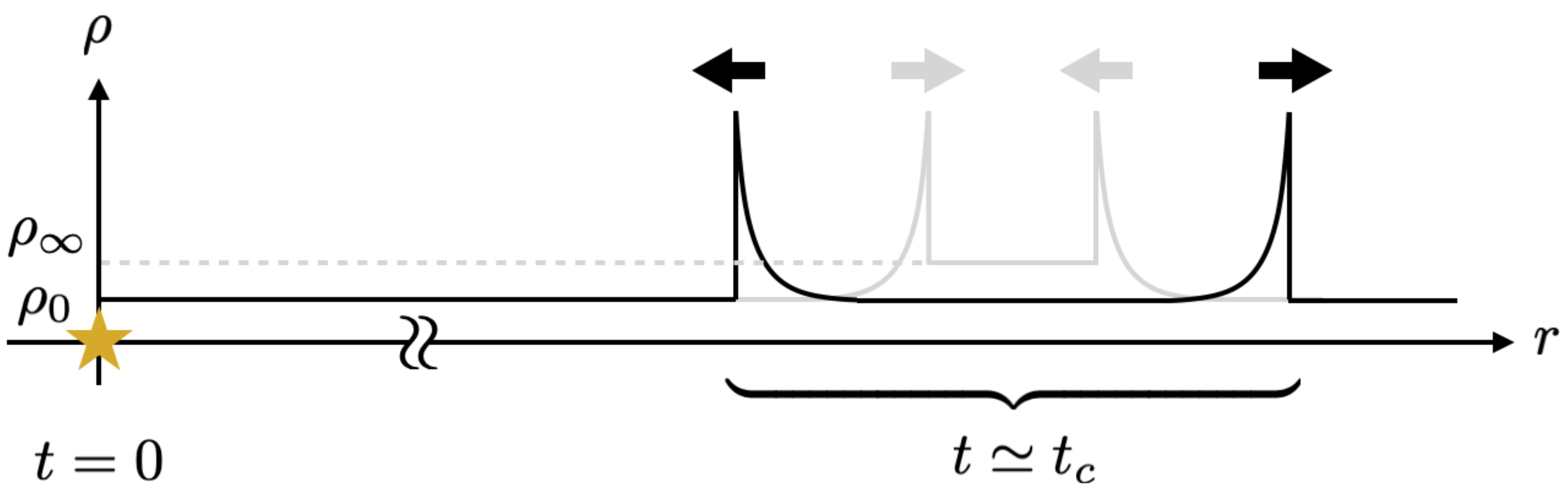} 
\caption{\small
After nucleation (which we set to occur at $t = 0$) at the yellow star, 
bubble expands until the collision time ($t = t_c$).
The released energy mainly goes to the fluid profile shown in gray.
The fluid energy densities inside and outside the expanding bubble 
are given by $\rho_0$ and $\rho_\infty$, respectively.
After the collision, the fluid is pushed into the broken phase of another bubble.
}
\label{fig:InitialCondition}
\end{center}
\end{figure}

Now let us state the problems we tackle in this paper more clearly.
Our initial relativistic fluid profile is a strong detonation.
Assuming that the first fluid collision -- which is simultaneous with the wall collision -- does not change the profile significantly,
we expect that this false-vacuum fluid would be pushed inside another expanding bubble, or the true vacuum.
Fig.~\ref{fig:InitialCondition} is an illustration of the setup.
Starting from this initial condition, we ask the following questions:
\begin{itemize}
\item
How does the relativistic fluid evolve in time?
\item
What is the implication to the GW production?
\end{itemize}
We tackle the former in Sec.~\ref{sec:Effective},
while we discuss the latter in Sec.~\ref{sec:Implications}
based on our findings in Sec.~\ref{sec:Effective}.

As briefly mentioned above, shock waves are one of the main sources of complication in the present system.
Generally a well-known phenomenon, they form when fluid is accelerated faster than the local speed of sound.
However, calculating their time evolution is difficult
because the shock front is a discontinuity at which derivatives diverge.
While numerical methods are available in the literature,
strong shock waves created in the phase transitions in the early universe are far beyond numerical tolerance.
Therefore, we develop an effective theory for the shock front in Sec.~\ref{sec:Effective}
that allows us to describe its time evolution in the relativistic limit. We then
compare its result with that of numerical simulations in an intermediate regime (see Fig.~\ref{fig:Relativisticity}).
Lastly, using the effective theory, 
we estimate how long such shock waves can last to source gravitational waves as thin, spherical objects before turning into sound waves or turbulence.

\section{Effective theory for relativistic shock propagation}
\label{sec:Effective}
\setcounter{equation}{0}

\subsection{Assumptions and approximations}

In this section, we model the system with a few quantities and derive the governing equations for propagation.
As mentioned in Sec.~\ref{sec:Review}, we consider the perfect-fluid form of the energy-momentum tensor:
\begin{align}
T_{\mu \nu}
&=
w u_\mu u_\nu - p g_{\mu \nu}.
\end{align}
Note that we have omitted and will omit the label ``plasma" in the following
because the system consists of fluid only after bubble collision.
We consider the equation of state
\begin{align}
p
&=
\frac{\rho}{3}
\label{eq:EOS}
\end{align}
to simplify the analysis.
This gives the speed of sound
\begin{align}
c_s^2
&=
\frac{1}{3}.
\end{align}
The property of the fluid can change depending on the microphysics,
but we expect that it does not change our qualitative results significantly.

Before moving on, we summarize our notations for subscripts here:
\begin{itemize}
\item
``wall":
quantities pertaining to the bubble wall.
For example, $\gamma_{\rm wall}$ is the $\gamma$ factor of the bubble wall.

\item
``max":
fluid quantities at the wall position during bubble expansion.
For example, $\rho_{\rm max}$ and $\gamma_{\rm max}$ are the fluid energy density and $\gamma$ factor at the wall position during bubble expansion.

\item
``peak":
fluid quantities at the shock front after bubble collision.
For example, $\rho_{\rm peak}$ and $\gamma_{\rm peak}$ are the fluid energy density and $\gamma$ factor
at the shock front after bubbles have collided and shock profiles are pushed into the true vacuum fluid.

\end{itemize}

\begin{figure}
\begin{center}
\includegraphics[width=0.75\columnwidth]{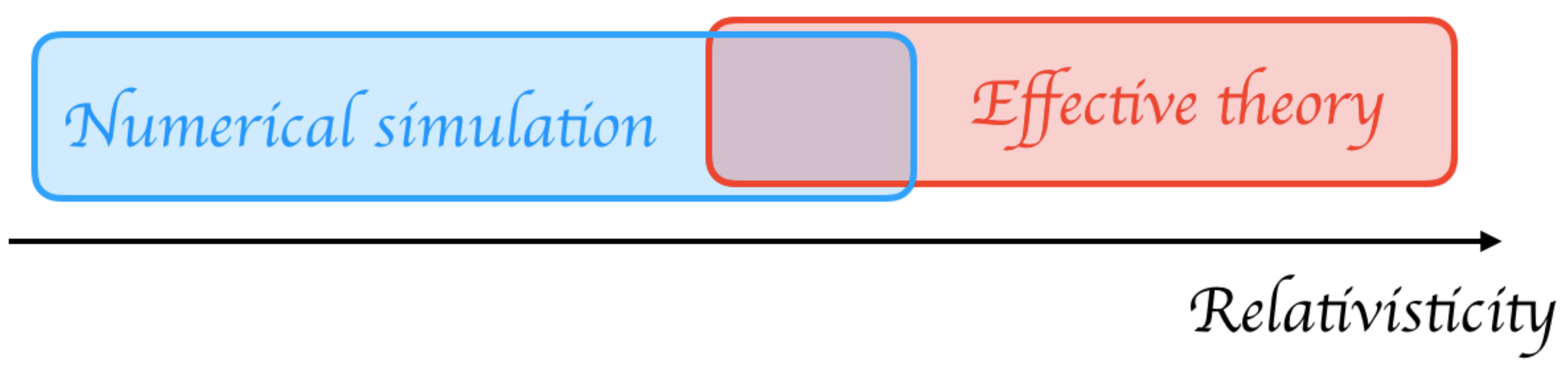} 
\caption{\small
Our strategy in this paper.
We construct an effective theory which works well with high relativisticity and then check its validity with the results from numerical simulations in the overlapping parameter space.
}
\label{fig:Relativisticity}
\end{center}
\end{figure}

\subsection{Equations of motion and Riemann invariants}
\label{subsec:EOM}

The evolution equations for the system simply come from the conservation of the energy-momentum tensor.
However, due to the discontinuities in the fluid profile,
an ambiguity arises when we write down the evolution equations in differential forms.
As we will see later, we have to impose the Rankine-Hugoniot relations to keep the conserved quantities intact across the discontinuities.

\subsubsection{Equations of motion}

Let us start from $\partial_\mu T^{\mu \nu} = 0$.
In the following, we assume $d$-dimensional spherical symmetry
(e.g. $d = 1$ is planar, $d = 2$ is cylindrical, and $d = 3$ is spherical).
Our final interest is the spherical case,
but we keep $d$ as a free parameter since we will check the validity of our theory using numerical simulation for the $d = 1$ case later.
The evolution equations reduce to
(see e.g. Ref.~\cite{McKee_1973})
\begin{align}
\partial_t u + \partial_r f + g
&= 0,
\end{align}
where
\begin{align}
u
&= 
\left(
\begin{matrix}
u_1 \\
u_2
\end{matrix}
\right)
=
\left(
\begin{matrix}
\displaystyle\frac{\rho + p v^2}{1 - v^2} \\[0.4cm]
\displaystyle\frac{(\rho + p) v}{1 - v^2} 
\end{matrix}
\right),
~~
f
= 
\left(
\begin{matrix}
\displaystyle\frac{(\rho + p) v}{1 - v^2} \\[0.4cm]
\displaystyle\frac{\rho v^2 + p}{1 - v^2}
\end{matrix}
\right),
~~
g
= 
\frac{d-1}{r}
\left(
\begin{matrix}
\displaystyle\frac{(\rho + p) v}{1 - v^2} \\[0.4cm]
\displaystyle\frac{(\rho + p) v^2}{1 - v^2}
\end{matrix}
\right).
\end{align}
In terms of $\rho$ and $v$, we have
\begin{align}
\partial_t
\left(
\begin{matrix}
\rho \\
v
\end{matrix}
\right)
 + A~\partial_r
\left(
\begin{matrix}
\rho \\
v
\end{matrix}
\right)
+ h
&= 0,
\label{eq:delrhov}
\end{align}
where $A$ and $h$ are
\begin{align}
A
&= 
\frac{1}{1 - c_s^2 v^2}
\left(
\begin{matrix}
(1 - c_s^2) v & \rho + p
\\[0.4cm]
\displaystyle \frac{c_s^2 (1 - v^2)^2}{\rho + p} & (1 - c_s^2) v
\end{matrix}
\right),
~~~~
h
=
\frac{d - 1}{r}
\left(
\begin{matrix}
\displaystyle
\frac{(\rho + p) v}{1 - c_s^2 v^2}
\\[0.4cm]
\displaystyle
- \frac{c_s^2 v^2 (1 - v^2)}{1 - c_s^2 v^2}
\end{matrix}
\right).
\end{align}

\subsubsection{Riemann invariants}

It is known that two conserved quantities exist along the eigenvalues
\begin{align}
v_{C_\pm}
&= 
\frac{v \pm c_s}{1 \pm v c_s}
\end{align}
of the matrix $A$.
Indeed, Eq.~(\ref{eq:delrhov}) can be rewritten as
\begin{align}
\pm
\frac{c_s}{\rho + p}
\left(
\frac{\partial}{\partial t}
+
v_{C_\pm}
\frac{\partial}{\partial r}
\right)
\rho
+
\frac{1}{1 - v^2}
\left(
\frac{\partial}{\partial t}
+
v_{C_\pm}
\frac{\partial}{\partial r}
\right)
v
\pm
\frac{v c_s}{1 \pm v c_s}
\frac{d - 1}{r}
&=
0.
\label{eq:delRpm}
\end{align}
This implies that, for $d = 1$, the quantities called the Riemann invariants
\begin{align}
R_\pm
&\equiv
\pm
\int d\rho~
\frac{c_s}{\rho + p}
+
\frac{1}{2} \ln \frac{1 + v}{1 - v},
\end{align}
are conserved along the directions defined by
\begin{align}
\left.
\frac{d}{dt}
\right|_\pm
&\equiv
\frac{\partial}{\partial t}
+
v_{C_\pm}
\frac{\partial}{\partial r}.
\end{align}
Note that these directions are paths of sound wave propagations
since $v_{C_\pm} = (v \pm c_s) / (1 \pm v c_s)$ are just the relativistic sums of $v$ and $c_s$.
They are called the forward and backward characteristics, respectively, for which we have used the label $C_\pm$.
For the equation of state (\ref{eq:EOS}), the conserved quantities become
\begin{align}
R_\pm
&=
\frac{1}{2} \ln
\left(
\rho^{\pm \sqrt{3}/2}~\frac{1 + v}{1 - v}
\right)
\simeq
\frac{1}{2} \ln
\left(
4 \rho^{\pm \sqrt{3}/2}~\gamma^2
\right),
\end{align}
where we took the relativistic limit in the last equality.

\subsection{Effective theory for relativistic shock propagation}

\subsubsection{Fundamental variables}
\label{subsubsec:Variables}

In this section we derive the effective theory for the shock wave propagation.
However, before doing so, let us talk about result of a numerical simulation for $d = 1$ with $(\gamma_{\rm wall}, \alpha) = (10,10)$ (Fig.~\ref{fig:Propagation}).
Note that the setup in this figure is not of our final interest
because the initial profile is obtained from an expanding {\it spherical} bubble
while the evolution is calculated with the assumption of {\it planar} symmetry.
The aim of this figure is to get a rough idea of the shock propagation after bubble collision
and to check the validity of the effective theory.
The length scale of the horizontal axis is chosen so that the collision time $t_c$ is unity
(with the nucleation time set to be zero as mentioned above).
We observe that
\begin{itemize}
\item
The initial position of the fluid front is $r/t_c \simeq 1$ because the wall is highly relativistic.
\item
Soon after the fluid is pushed into the low-energy plasma,
the peak values of $\rho$ and $\gamma^2$ get rearranged to certain values.
After that, the peak values evolve gradually.
\item
At the same time, a discontinuity detaches from the peak and evolves toward the rear side of the fluid profile.
\end{itemize}
As far as the GW production is concerned, we are not much interested in the tail part of the fluid profile.
Rather, we focus on the time evolution of the high-energy peak where the energy-momentum tensor takes large values.
Therefore, we model the strong shock front using the following five variables:
\begin{itemize}
\item
$\gamma$-factor of the shock wave
\begin{align}
\gamma^2_s (t)
&\equiv
\frac{1}{1 - (dr_s(t)/dt)^2},
\end{align}

\item
Peak values of the fluid energy density and their $\gamma$-factors
\begin{align}
\rho_{\rm peak} (t)
&\equiv 
\rho (t, r = r_s (t)),
~~~~
\gamma^2_{\rm peak} (t)
\equiv 
\gamma^2 (t, r = r_s (t)),
\end{align}

\item
Derivatives of the fluid energy density and the $\gamma$-factor at the shock front
\begin{align}
\ln\rho' (t)
&\equiv 
\left.
\frac{\partial \ln \rho}{\partial r}
\right|_{r = r_s(t)}
\left(
= 
\frac{\left. d \rho / dr \right|_{r = r_s(t)}}{\rho_{\rm peak} (t)}
\right),
\\[0.2cm]
\ln\gamma^{2'} (t)
&\equiv 
\left.
\frac{\partial \ln (\gamma^2)}{\partial r}
\right|_{r = r_s(t)}
\left(
=
\frac{\left. d (\gamma^2) / dr \right|_{r = r_s(t)}}{\gamma^2_{\rm peak} (t)}
\right),
\end{align}
\end{itemize}
where $r_s (t)$ is the position of the shock front. 
In the following, we derive the governing equations for these variables.

\subsubsection{Equations}
\label{subsubsec:Equations}

\paragraph*{Rankine-Hugoniot relations.}

As discussed in Sec.~\ref{subsec:EOM}, 
equations of motion in differential forms pose an ambiguity due to the discontinuity in the fluid profile.
Therefore, we resort to the Rankine-Hugoniot conditions which guarantee the conservation law on both sides of the shock.
In the present system, we have two Rankine-Hugoniot conditions arising from two-component
equations of motion.
They can be read off as (see e.g. Ref.~\cite{McKee_1973}),
\begin{align}
p_{\rm peak}
= 
\frac{p_0 + \rho_0 v_{\rm peak} v_s}{1 - v_{\rm peak}v_s},
~~~~
v_s
=
\frac{(p_{\rm peak} + \rho_{\rm peak}) v_{\rm peak}}
{p_{\rm peak} v_{\rm peak}^2 + \rho_{\rm peak} - \rho_0 (1 - v_{\rm peak}^2)}.
\end{align}
Assuming the equation of state (\ref{eq:EOS}), these reduce to
\begin{align}
\gamma_s^2
&= 
\gamma_{\rm peak}^2
\left(
\displaystyle 1 + \frac{1}{2\gamma_{\rm peak}^2}
+
\sqrt{1 - \frac{1}{\gamma_{\rm peak}^2}}
\sqrt{1 - \frac{1}{4\gamma_{\rm peak}^2}}
\right),
\\
\frac{\rho_{\rm peak}}{\rho_0}
&= 
1 
+ 
\frac{8}{3}
\gamma_{\rm peak}^2
\sqrt{1 - \frac{1}{\gamma_{\rm peak}^2}}
\left(
\sqrt{1 - \frac{1}{\gamma_{\rm peak}^2}}
+
\sqrt{1 - \frac{1}{4\gamma_{\rm peak}^2}}
\right).
\end{align}
They further get simplified in the relativistic limit:
\begin{align}
\gamma_s^2
&= 
2 \gamma^2_{\rm peak},
~~~~
\displaystyle
\frac{\rho_{\rm peak}}{\rho_0}
= 
\displaystyle
\frac{16}{3}\gamma^2_{\rm peak}.
\label{eq:RH}
\end{align}
Note that these are constraint equations.

\paragraph*{Time evolution equations.}

\begin{figure}
\begin{center}
\includegraphics[width=\columnwidth]{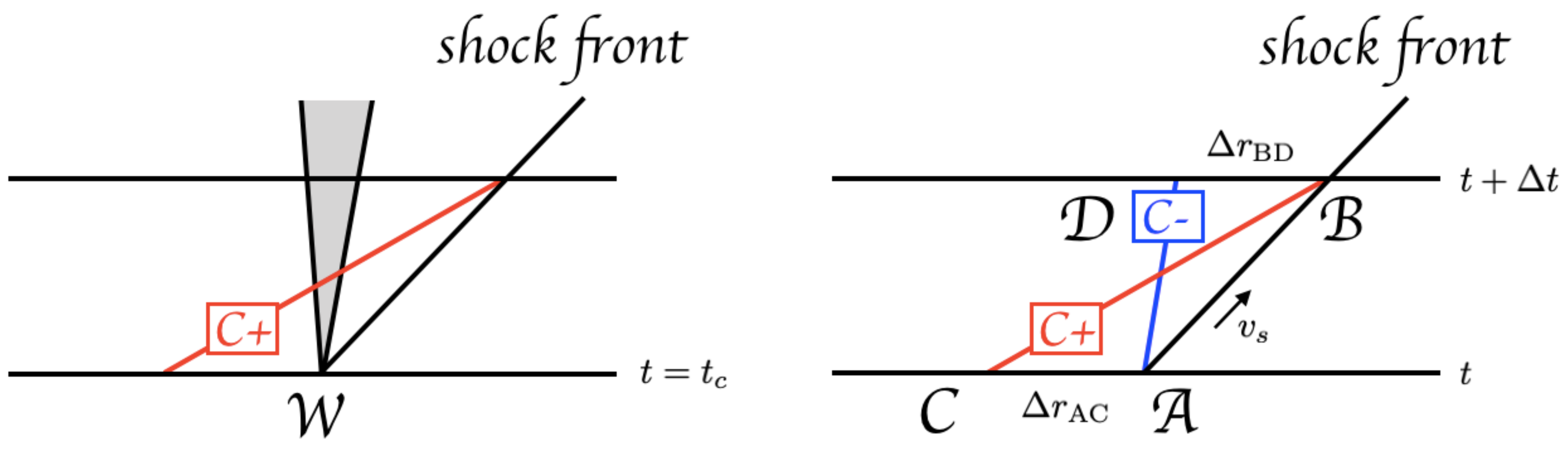}
\caption{ \small
Characteristics around the shock wave.
(Left) How to obtain initial conditions 
$(\gamma_s^2, \rho_{\rm peak}, \gamma_{\rm peak}^2, \ln \rho', \ln \gamma^{2'})$ around $t = t_c$
using the characteristic $C_+$.
(Right) How to obtain evolution equations using the characteristics $C_+$ and $C_-$.
}
\label{fig:Characteristic}
\end{center}
\end{figure}

In order to derive the time evolution equations, we use the time evolution of 
the Riemann invariants (Eq.~(\ref{eq:delRpm})) along the two characteristics $C_\pm$.
Fig.~\ref{fig:Characteristic} is a zoom-in around the shock front
with the red and blue lines describing the direction of the characteristics.
The rough idea is to relate the time evolution of the peak values 
(A $\to$ B) with the spatial derivative of the profile (C $\to$ A or D $\to$ B)
by connecting B and C or A and D with the characteristics.
We take the relativistic limit in the following.

Let us first use $C_+$. By considering A, B, and C, we have
\begin{align}
&\Delta t
~
\partial_t
\left[
\frac{1}{2} \ln \left( 4 \rho_{\rm peak}^{\sqrt{3}/2} \gamma_{\rm peak}^2 \right)
\right]
\nonumber \\
&=
\left.
\frac{1}{2} \ln \left( 4 \rho^{\sqrt{3}/2} \gamma^2 \right)
\right|_{\rm B}
-
\left.
\frac{1}{2} \ln \left( 4 \rho^{\sqrt{3}/2} \gamma^2 \right)
\right|_{\rm A}
\nonumber \\
&=
\left.
\frac{1}{2} \ln \left( 4 \rho^{\sqrt{3}/2} \gamma^2 \right)
\right|_{\rm C}
-
\left.
\frac{1}{2} \ln \left( 4 \rho^{\sqrt{3}/2} \gamma^2 \right)
\right|_{\rm A}
-
\Delta t~
\frac{v_{\rm peak} c_s}{1 + v_{\rm peak} c_s}
\frac{d - 1}{r_s (t)} 
\nonumber \\
&=
- \Delta r_{\rm AC}
~
\partial_r
\left.
\left[
\frac{1}{2} \ln \left( 4 \rho^{\sqrt{3}/2} \gamma^2 \right)
\right]
\right|_{\rm peak}
-
\Delta t~
\frac{\sqrt{3} - 1}{2}
\frac{d - 1}{t},
\label{eq:C+Derivation}
\end{align}
where we used the time evolution of $R_+$ (Eq.~(\ref{eq:delRpm})) from the second to third line.
Since $C_+$, $C_-$, fluid motion and shock front are all extremely inclined to the forward direction in the relativistic limit
(see Fig.~\ref{fig:Characteristic}), we also used $v_{\rm peak} \simeq 1$.
Furthermore, we replaced $r_s$ with $t$ in the last line by taking the relativistic limit.
The relation between $\Delta t$ and $\Delta r_{\rm AC}$ is obtained from Eq.~(\ref{eq:RH}) and
\begin{align}
\left(
v_{C_+}|_{\rm peak}
- 
v_s
\right)
\Delta t
&= 
\Delta r_{\rm AC},
\end{align}
as
\begin{align}
\Delta r_{\rm AC}
&= 
\frac{2\sqrt{3} - 3}{4} \frac{\Delta t}{\gamma_{\rm peak}^2}.
\end{align}
Therefore, we get
\begin{align}
\frac{\sqrt{3}}{2}
\partial_t \ln \rho_{\rm peak} 
+
\partial_t \ln \gamma_{\rm peak}^2
&=
-
\frac{2\sqrt{3} - 3}{4} \frac{1}{\gamma_{\rm peak}^2}
\left[
\frac{\sqrt{3}}{2} \ln \rho'
+ 
\ln \gamma^{2'}
\right]
-
\frac{(\sqrt{3} - 1)(d - 1)}{t}.
\label{eq:C+}
\end{align}
The other evolution is obtained from the consideration of A, B, and D
with A and D connected by $C_-$:
\begin{align}
&\Delta t
~
\partial_t
\left[
\frac{1}{2} \ln \left( 4 \rho_{\rm peak}^{- \sqrt{3}/2} \gamma_{\rm peak}^2 \right)
\right]
\nonumber \\
&=
\left.
\frac{1}{2} \ln \left( 4 \rho^{- \sqrt{3}/2} \gamma^2 \right)
\right|_{\rm B}
-
\left.
\frac{1}{2} \ln \left( 4 \rho^{- \sqrt{3}/2} \gamma^2 \right)
\right|_{\rm A}
\nonumber \\
&=
\left.
\frac{1}{2} \ln \left( 4 \rho^{- \sqrt{3}/2} \gamma^2 \right)
\right|_{\rm B}
-
\left.
\frac{1}{2} \ln \left( 4 \rho^{- \sqrt{3}/2} \gamma^2 \right)
\right|_{\rm D}
+
\Delta t~
\frac{v_{\rm peak} c_s}{1 - v_{\rm peak} c_s}
\frac{d - 1}{r_s (t)} 
\nonumber \\
&=
\Delta r_{\rm BD}
~
\partial_r
\left.
\left[
\frac{1}{2} \ln \left( 4 \rho^{- \sqrt{3}/2} \gamma^2 \right)
\right]
\right|_{\rm peak}
+
\Delta t~
\frac{\sqrt{3} + 1}{2}
\frac{d - 1}{t},
\label{eq:C-Derivation}
\end{align}
where we used the time evolution of $R_-$ (Eq.~(\ref{eq:delRpm})) from the second to third line
and $v_{\rm peak} \simeq 1$ from the third to fourth line.
The relation between $\Delta t$ and $\Delta r_{\rm BD}$ is calculated from Eq.~(\ref{eq:RH}) and
\begin{align}
\left(
v_s
-
v_{C_-}|_{\rm peak}
\right)
\Delta t
&= 
\Delta r_{\rm BD},
\end{align}
as
\begin{align}
\Delta r_{\rm BD}
&= 
\frac{2\sqrt{3} + 3}{4} \frac{\Delta t}{\gamma_{\rm peak}^2}.
\end{align}
The result is
\begin{align}
- \frac{\sqrt{3}}{2}
\partial_t \ln \rho_{\rm peak}
+
\partial_t \ln \gamma_{\rm peak}^2
&=
\frac{2\sqrt{3} + 3}{4} \frac{1}{\gamma_{\rm peak}^2}
\left[
- \frac{\sqrt{3}}{2} \ln \rho'
+ 
\ln \gamma^{2'}
\right]
+
\frac{(\sqrt{3} + 1)(d - 1)}{t}.
\label{eq:C-}
\end{align}
Here, we notice one interesting thing.
Since $\rho_{\rm peak}$ is related to $\gamma_{\rm peak}^2$ via Eq.~(\ref{eq:RH}),
$\ln \rho'$ must be related to $\ln \gamma^{2'}$ to make Eqs.~(\ref{eq:C+}) and (\ref{eq:C-}) consistent.
After some calculations, we find that
\begin{align}
\ln \rho'
&=
\frac{7}{6}
\ln \gamma^{2'}
+
\frac{2(d - 1) \gamma_{\rm peak}^2}{t}.
\label{eq:C+C-1}
\end{align}
Substituting this into either Eq.~(\ref{eq:C+}) or (\ref{eq:C-}) and using Eq.~(\ref{eq:RH}),
we obtain
\begin{align}
\partial_t 
\gamma_{\rm peak}^2
&=
- \frac{1}{8} \ln \gamma^{2'}
-
\frac{1}{2} \frac{(d - 1) \gamma_{\rm peak}^2}{t}.
\label{eq:C+C-2}
\end{align}

\paragraph*{Energy and momentum domination by shock front.}

So far, we have derived only four constraint or evolution equations for five variables.
The last condition we add is the energy and momentum dominance by the shock front.
In order to do so, we need a specific ansatz for the fluid profile.
In our analysis, we adopt
\begin{align}
\rho
&= 
\rho_{\rm peak} e^{\ln\rho' (r - r_s)},
~~
\gamma^2
= 
\gamma^2_{\rm peak} e^{\ln \gamma^{2'} (r - r_s)}.
\end{align}
Note that these profiles are only estimates and the choice is rather arbitrary.
However, as long as the energy and momentum of the system are dominated by the shocks,
we expect that the relation ``(energy or momentum at the peak) $\times$ (thickness of the fluid) 
$\simeq$ const." holds well.
Indeed, in the relativistic limit, the energy or momentum calculated from the above ansatz 
(for $d = 2$ and $3$, energy or momentum per unit angle and solid angle, respectively)
becomes
\begin{align}
\sigma
&\simeq
\left\{
\begin{matrix}
1 \\
t \\
t^2
\end{matrix}
\right\}
\times
\displaystyle
\int dr~
\frac{4}{3} \rho \gamma^2
= 
\left\{
\begin{matrix}
1 \\
t \\
t^2
\end{matrix}
\right\}
\times
\frac{4}{3}~\frac{\rho_{\rm peak} \gamma_{\rm peak}^2}{\ln \rho' + \ln \gamma^{2'}}
~~~~
{\rm for}
~~~~
\left\{
\begin{matrix}
d = 1 \\[0.08cm]
d = 2 \\[0.08cm]
d = 3
\end{matrix}
\right\},
\label{eq:sigma}
\end{align}
and satisfies (energy or momentum at the peak) $ = \rho_{\rm peak} \gamma_{\rm peak}^2$
and (thickness of the fluid) $ = 1 / (\ln \rho' + \ln \gamma^{2'})$.
Note that $\sigma$ has different mass dimensions depending on the value of $d$.
We use Eq.~(\ref{eq:sigma}) with $\sigma$ calculated from the initial profile
as the last ingredient of our effective theory.
Though this constraint is rather qualitative in contrast to the other four equations, 
our effective theory (\ref{eq:RH}), (\ref{eq:C+}), (\ref{eq:C-}), and (\ref{eq:sigma}) 
describes the duration of the shock rather correctly as we will see later.

\subsubsection{Solution}
\label{subsubsec:Solution}

The five equations (\ref{eq:RH}), (\ref{eq:C+}), (\ref{eq:C-}), and (\ref{eq:sigma}) can be solved analytically. 
For $d = 1$, the solution becomes
\begin{align}
\frac{1}{\gamma_{\rm s}^2 (t)}
&=
\frac{8}{39} \left( \frac{\rho_0}{\sigma} \right) (t - t_c) + \frac{1}{\gamma_{\rm s}^2 (t_c)},
\end{align}
\begin{align}
\frac{\rho_0}{\rho_{\rm peak} (t)}
&=
\frac{1}{13} \left( \frac{\rho_0}{\sigma} \right) (t - t_c) + \frac{\rho_0}{\rho_{\rm peak} (t_c)},
~~~~
\frac{1}{\gamma_{\rm peak}^2 (t)}
=
\frac{16}{39} \left( \frac{\rho_0}{\sigma} \right) (t - t_c) + \frac{1}{\gamma_{\rm peak}^2 (t_c)},
\end{align}
\begin{align}
\ln \rho' (t)
&=
\frac{448}{117} \left( \frac{\rho_0}{\sigma} \right) \gamma_{\rm peak}^4(t),
~~~~
\ln {\gamma^2}' (t)
=
\frac{128}{39} \left( \frac{\rho_0}{\sigma} \right) \gamma_{\rm peak}^4(t),
\end{align}
with the following constraints among the initial values
\begin{align}
\gamma_s^2 (t_c)
= 
2\gamma^2_{\rm peak} (t_c),
&~~~~
\frac{\rho_{\rm peak} (t_c)}{\rho_0}
= 
\frac{16}{3} \gamma^2_{\rm peak} (t_c),
\nonumber \\
\ln \rho' (t_c)
= 
\frac{448}{117} 
\left(
\frac{\rho_0}{\sigma} 
\right)
\gamma^4_{\rm peak} (t_c),
&~~~~
\ln \gamma^{2'} (t_c)
= 
\frac{128}{39} 
\left(
\frac{\rho_0}{\sigma} 
\right)
\gamma^4_{\rm peak} (t_c).
\label{eq:IC_d=1}
\end{align}
On the other hand, for $d = 3$, the solution becomes
\begin{align}
\frac{1}{\gamma_s^2 (t)}
&=
\frac{8}{87} 
\left(
\frac{\rho_0}{\sigma}
\right)
\left[
t^3
-
\left(
\frac{t}{t_c}
\right)^\delta
t_c^3
\right]
+
\frac{1}{\gamma_{\rm s}^2 (t_c)}
\left(
\frac{t}{t_c}
\right)^\delta,
\\
\frac{\rho_0}{\rho_{\rm peak} (t)}
&=
\frac{1}{29} 
\left(
\frac{\rho_0}{\sigma}
\right)
\left[
t^3
-
\left(
\frac{t}{t_c}
\right)^\delta
t_c^3
\right]
+
\frac{\rho_0}{\rho_{\rm peak} (t_c)}
\left(
\frac{t}{t_c}
\right)^\delta,
\\
\frac{1}{\gamma_{\rm peak}^2 (t)}
&=
\frac{16}{87} 
\left(
\frac{\rho_0}{\sigma}
\right)
\left[
t^3
-
\left(
\frac{t}{t_c}
\right)^\delta
t_c^3
\right]
+
\frac{1}{\gamma_{\rm peak}^2 (t_c)}
\left(
\frac{t}{t_c}
\right)^\delta,
\end{align}
\begin{align}
\ln \rho' (t)
&=
\frac{448}{117} \left( \frac{\rho_0}{\sigma} \right) t^2 \gamma_{\rm peak}^4(t)
+
\frac{24}{13} \frac{\gamma_{\rm peak}^2(t)}{t},
~~~~
\ln {\gamma^2}' (t)
=
\frac{128}{39} \left( \frac{\rho_0}{\sigma} \right) t^2 \gamma_{\rm peak}^4(t)
- 
\frac{24}{13} \frac{\gamma_{\rm peak}^2(t)}{t},
\end{align}
with the terms with the exponent $\delta \equiv 10/13$ becoming numerically negligible at late times
compared to other terms.
The constraints among the initial values are given by
\begin{align}
\gamma_s^2 (t_c)
= 
2\gamma^2_{\rm peak} (t_c),
&~~~~
\frac{\rho_{\rm peak} (t_c)}{\rho_0}
= 
\frac{16}{3} \gamma^2_{\rm peak} (t_c),
\nonumber \\
\ln \rho' (t_c)
= 
\frac{448}{117} \left( \frac{\rho_0}{\sigma} \right) t_c^2 \gamma_{\rm peak}^4(t_c)
+
\frac{24}{13} \frac{\gamma_{\rm peak}^2(t_c)}{t_c},
&~~~~
\ln \gamma^{2'} (t_c)
= 
\frac{128}{39} \left( \frac{\rho_0}{\sigma} \right) t_c^2 \gamma_{\rm peak}^4(t_c)
- 
\frac{24}{13} \frac{\gamma_{\rm peak}^2(t_c)}{t_c}.
\label{eq:IC_d=3}
\end{align}
This solution implies that the lifetime of the relativistic fluid profile is qualitatively given by
\begin{align}
\tau
&\equiv
\left( \frac{\sigma}{\rho_0} \right)^{1/3},
\label{eq:tau}
\end{align}
as long as we take only propagation effects into account.
Note that the thickness of the fluid ($\sim 1/\ln \rho'$ or $1/\ln {\gamma^2}'$) is also determined by this timescale as
\begin{align}
\frac{\rm (thickness~of~the~fluid~profile)}{\rm (size~of~the~expanding~fluid~bubble)}
&\sim
\frac{{1/\ln \rho'} {\rm ~or~} 1/\ln {\gamma^2}'}{t}
\sim
\left( \frac{\rho_0}{\sigma} \right) t^3
\sim 
\left( \frac{t}{\tau} \right)^3.
\end{align}
In Sec.~\ref{subsubsec:IC}, we discuss how to determine the initial conditions (\ref{eq:IC_d=1}) and (\ref{eq:IC_d=3}) 
just after bubble collision from the profile of expanding bubbles.

\subsubsection{Initial conditions}
\label{subsubsec:IC}

So far, we have discussed the effective theory without mentioning the initial conditions.
For detonations, which we have used in this paper, 
the initial values $\rho_{\rm peak}$ or $\gamma_{\rm peak}^2$ can be estimated by 
the Riemann invariant $R_+$ along the characteristic $C_+$
as shown in the left panel of Fig.~\ref{fig:Characteristic}.\footnote{
For hybrids, we cannot use the estimate (\ref{eq:IC}) 
because the shock does not start from the wall position $W$ in the left panel of Fig.~\ref{fig:Characteristic}.
However, in Sec.~\ref{sec:Implications}, we slightly extrapolate Eq.~(\ref{eq:IC}) to estimate the
GW production for the hybrid regime as well.
\label{fn:IC}
}
In the relativistic limit, the relation is given by
\begin{align}
\rho_{\rm max}^{\sqrt{3}/2} \gamma_{\rm max}^2
&=
\rho_{\rm peak}^{\sqrt{3}/2} (t_c) \gamma_{\rm peak}^2 (t_c).
\label{eq:IC}
\end{align}
With the Rankine-Hugoniot relation (\ref{eq:RH}), we have
\begin{align}
\frac{\rho_{\rm peak} (t_c)}{\rho_0}
&= 
\left(
\frac{3}{16}
\right)^{2 \sqrt{3} - 4}
\left(
\frac{\rho_{\rm max}}{\rho_0}
\right)^{2 \sqrt{3} - 3}
\gamma_{\rm max}^{- 4 \sqrt{3} + 8},
\\
\gamma_{\rm peak}^2 (t_c)
&= 
\left(
\frac{3}{16}
\right)^{2 \sqrt{3} - 3}
\left(
\frac{\rho_{\rm max}}{\rho_0}
\right)^{2 \sqrt{3} - 3}
\gamma_{\rm max}^{- 4 \sqrt{3} + 8}.
\end{align}
In Sec.~\ref{subsec:Comparison}, we indeed see that this relation gives a good estimate on the initial values.

\subsection{Comparison with numerical simulations}
\label{subsec:Comparison}

\begin{figure}
\begin{center}
\includegraphics[width=0.95\columnwidth]{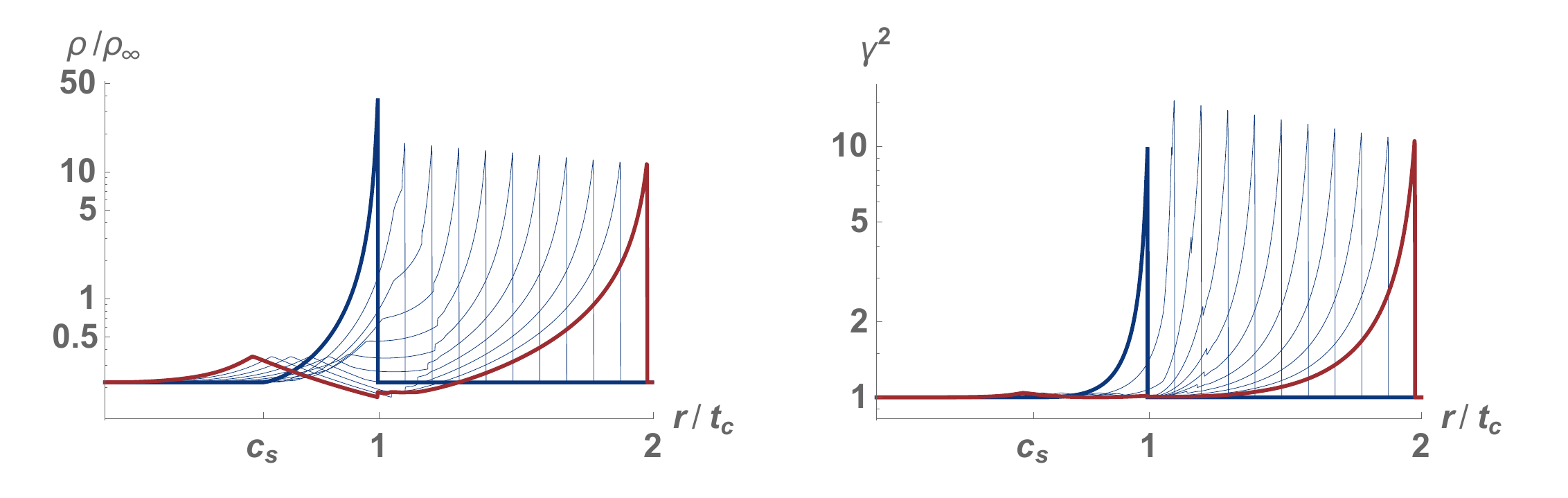}
\includegraphics[width=0.95\columnwidth]{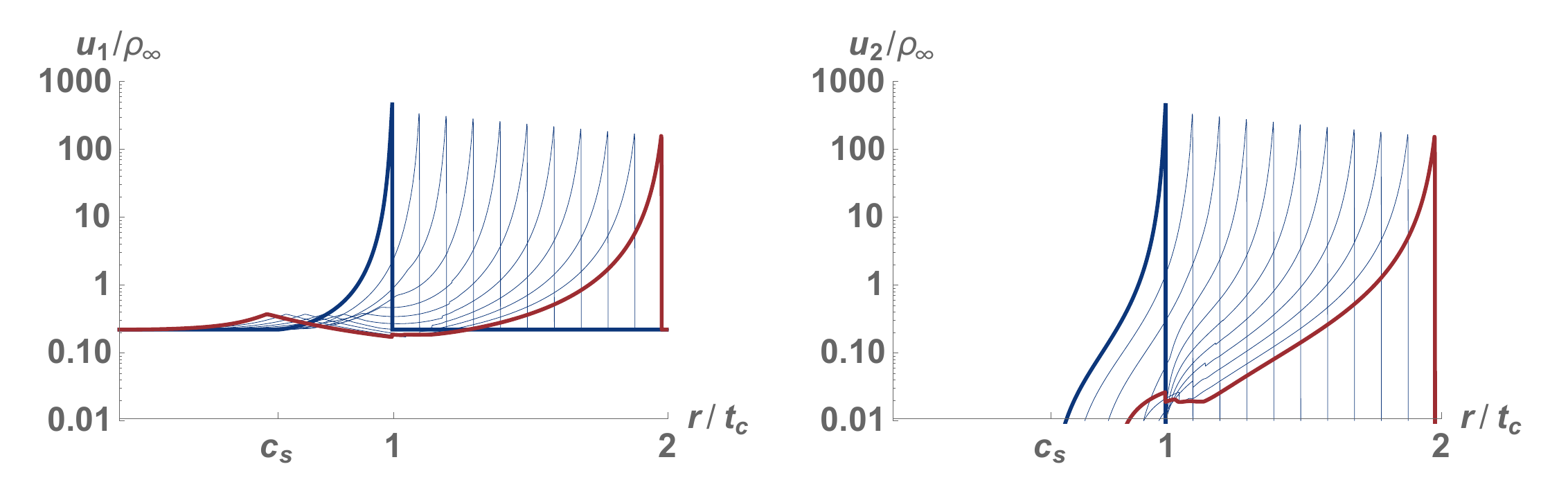}
\caption{\small
Numerical simulation results for the shock wave propagation for $(\gamma_{\rm wall}, \alpha) = (10, 10)$
(which translates to $\gamma_{\rm max}^2 \simeq 10$) for $d = 1$. 
See the main text for details.
}
\label{fig:Propagation}
\end{center}
\end{figure}

\begin{figure}
\begin{center}
\includegraphics[width=0.42\columnwidth]{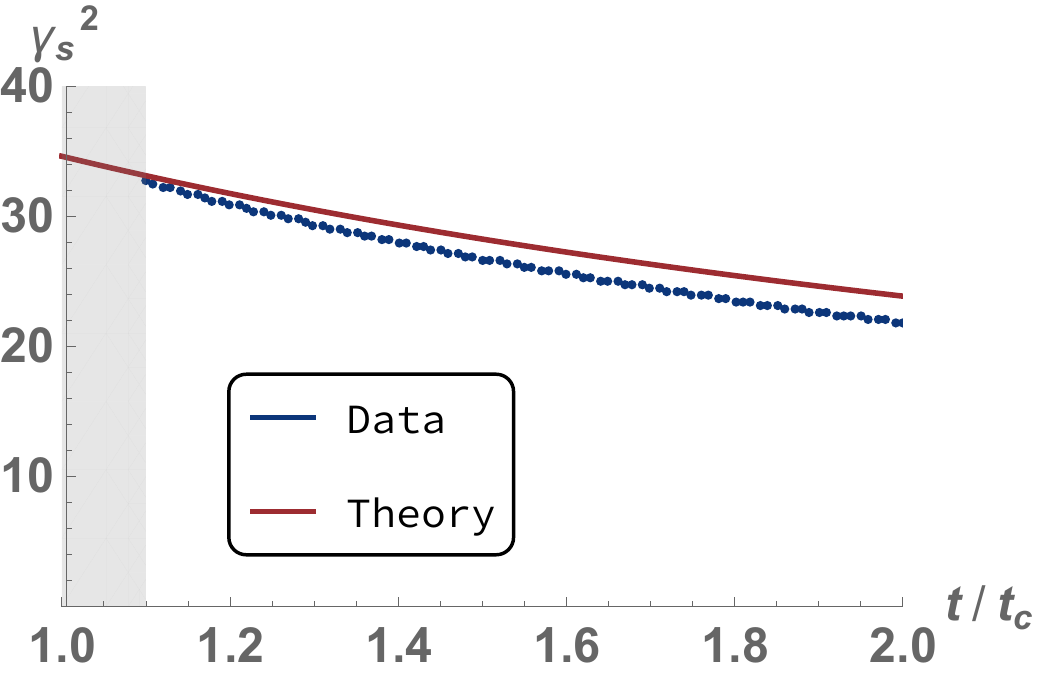}
\vskip 1cm
\includegraphics[width=0.42\columnwidth]{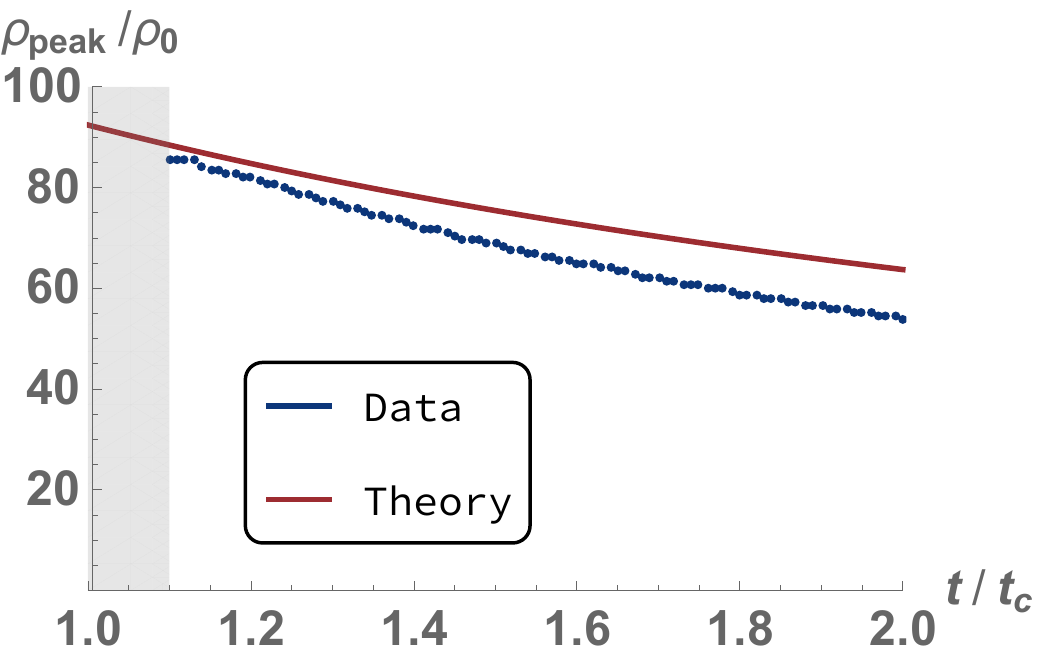}
\includegraphics[width=0.42\columnwidth]{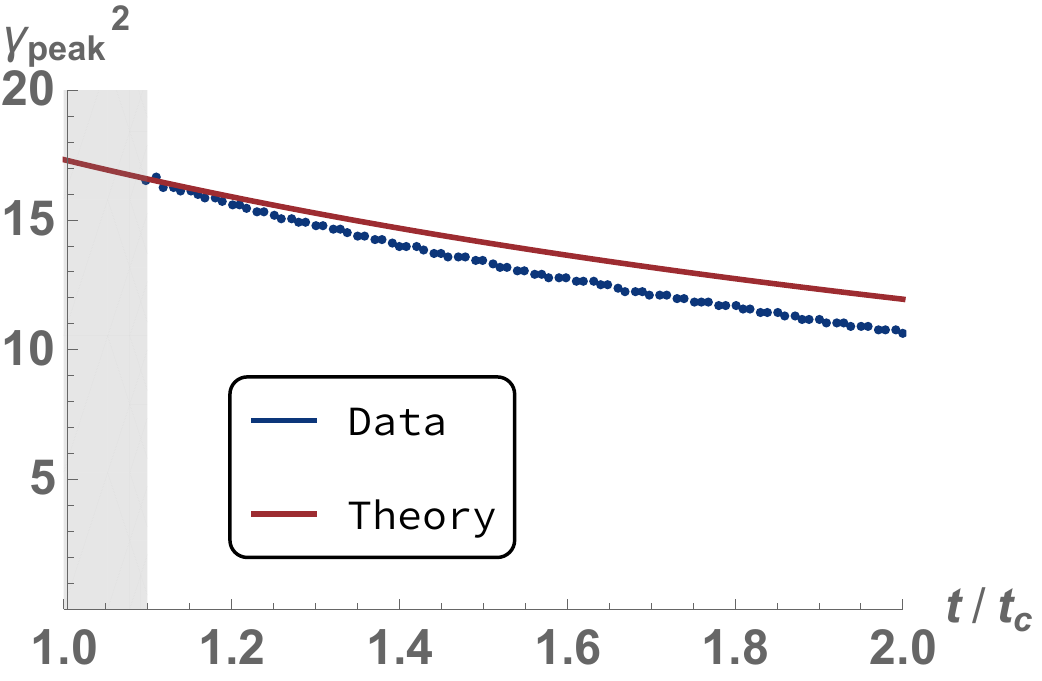}
\vskip 1cm
\includegraphics[width=0.42\columnwidth]{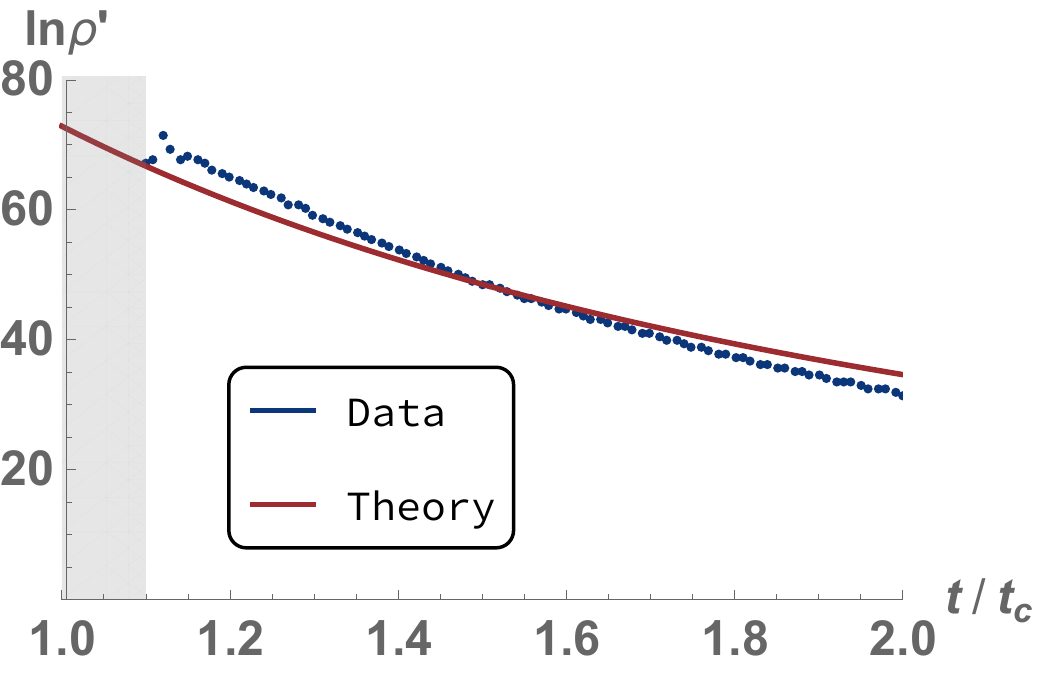}
\includegraphics[width=0.42\columnwidth]{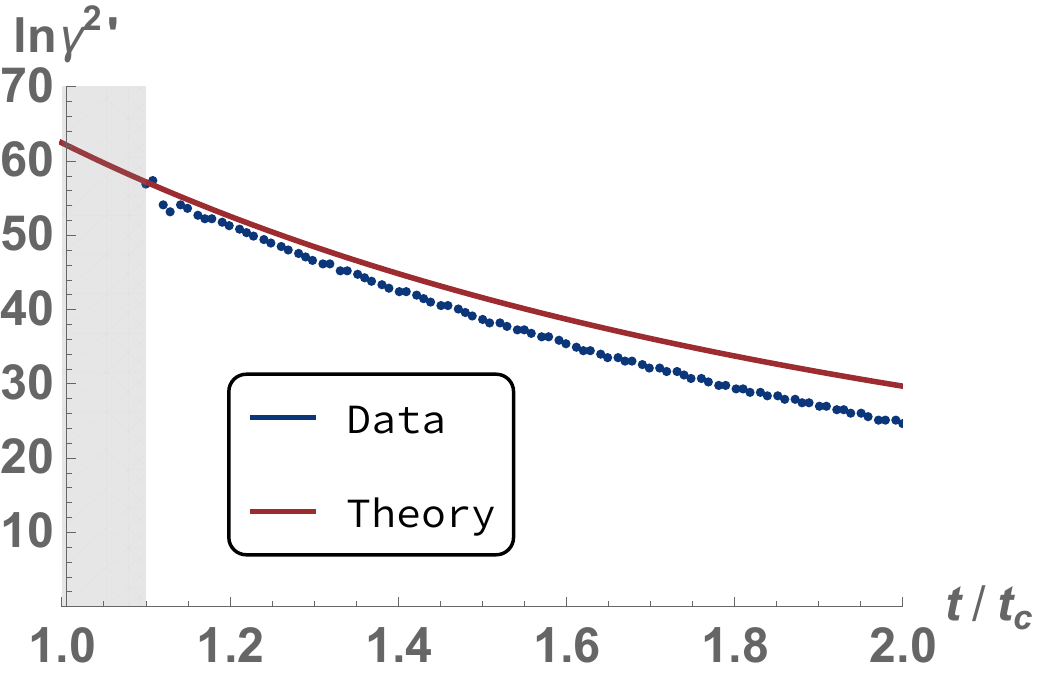}
\vskip 0.5cm
\caption{\small
Time evolution of the quantities in the effective theory for relativistic shock propagation.
(Blue) Numerical values estimated from the simulation.
(Red) Prediction from the effective theory.
}
\label{fig:Evolution}
\end{center}
\end{figure}

\begin{figure}
\begin{center}
\includegraphics[width=0.49\columnwidth]{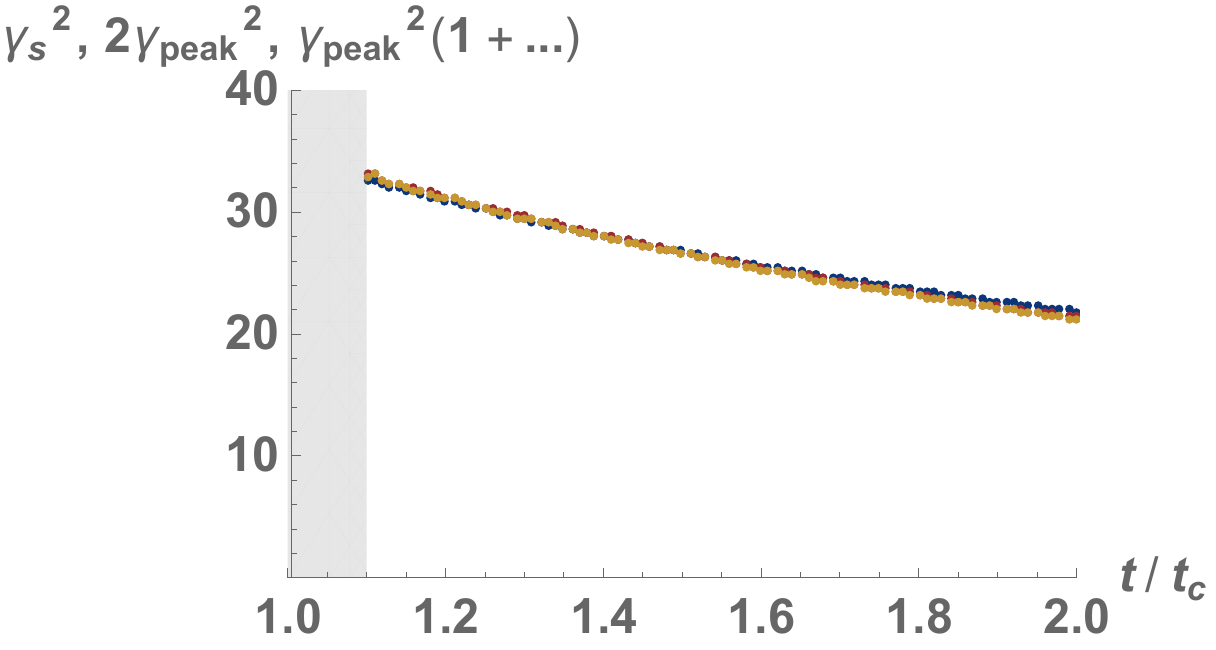}
\includegraphics[width=0.49\columnwidth]{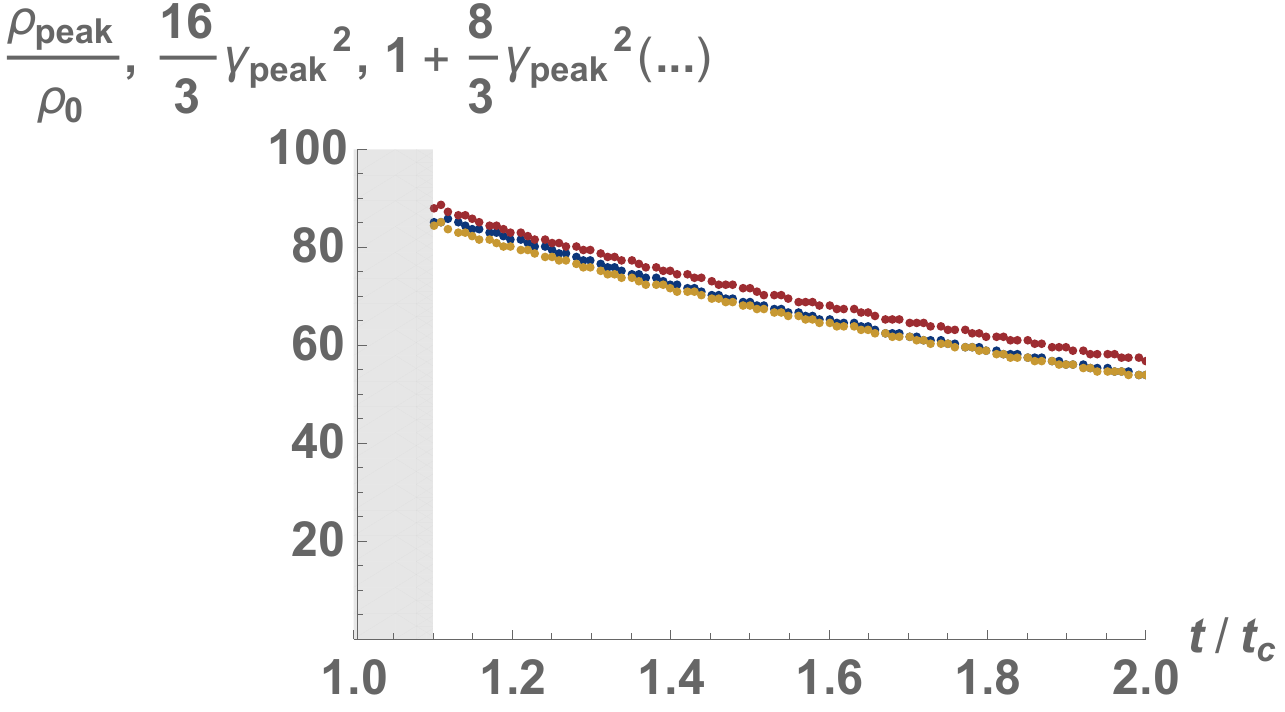}
\caption{\small
A numerical check for the Rankine-Hugoniot conditions (\ref{eq:RH}).
(Left) $\gamma_s^2$ (blue), $2\gamma_{\rm peak}^2$ (red), and Eq.~(\ref{eq:RH1Beyond}) (yellow).
(Right) $\rho_{\rm peak} / \rho_0$ (blue), $(16/3)\gamma_{\rm peak}^2$ (red), and Eq.~(\ref{eq:RH2Beyond}) (yellow).
}
\label{fig:RH}
\vskip 1cm
\includegraphics[width=0.45\columnwidth]{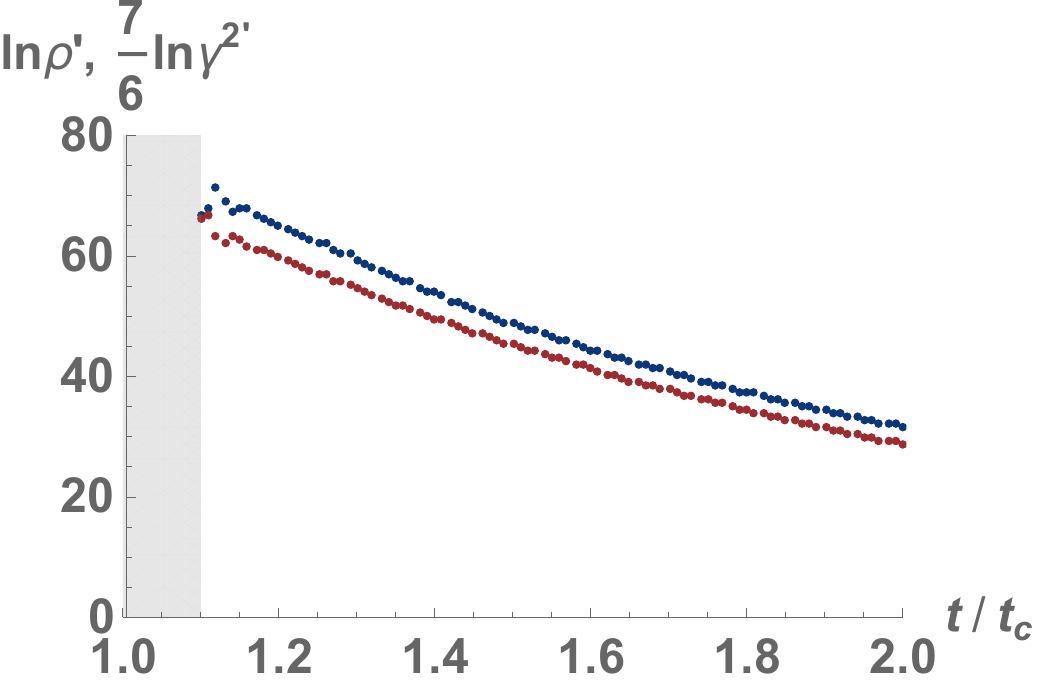}
\caption{\small
A numerical check for the condition (\ref{eq:C+C-1}) derived from 
the Rankine-Hugoniot conditions (\ref{eq:RH}) and the two evolution equations (\ref{eq:C+}) and (\ref{eq:C-}).
The points are  $\ln \rho'$ (blue) and $(7/6) \ln {\gamma^2}'$ (red).
}
\label{fig:EOM}
\end{center}
\end{figure}

Let us check the validity of our effective theory using numerical simulations with moderate relativisticity.
Adopting the numerical scheme proposed in Ref.~\cite{Kurganov_2000}, we perform a numerical simulation for $(\gamma_{\rm wall}, \alpha) = (10, 10)$,
values with which taking the relativistic limit is valid (see Fig.~\ref{fig:Relativisticity}).
The result is shown in Fig.~\ref{fig:Propagation}.
The initial profiles, denoted by thick blue lines, are calculated per Ref.~\cite{Espinosa:2010hh}, and we evolved the system from $t = t_c$ to $t = 2t_c$.
We observe several features explained in Sec.~\ref{subsubsec:Variables}.
From this fluid evolution, we extracted the five quantities introduced in Sec.~\ref{subsubsec:Variables} 
and plotted them in Fig.~\ref{fig:Evolution}.
Though the numerical scheme in Ref.~\cite{Kurganov_2000} tracks the shock evolution quite well,
numerical viscosity still plays a role.
Therefore, we obtain the peak quantities by extrapolation as explained in Appendix.~\ref{app:Detail}.
In Fig.~\ref{fig:Evolution}, the blue points are the quantities obtained by this extrapolation, 
while the red lines are the prediction from our effective theory.
For $t_c < t < 1.1 \times t_c$, the blue points are not shown because the peak structure is not clear during that time since the substructures have not completely detached from the peak (e.g. the top-left panel of Fig.~\ref{fig:Propagation}).
We see that our theory predicts the shock evolution qualitatively well, 
while it does not coincide with the data perfectly.
There are two possible reasons for this deviation:
\begin{itemize}
\item
The Rankine-Hugoniot relations (\ref{eq:RH}) and the time evolution equations (\ref{eq:C+}) and (\ref{eq:C-}) 
are derived in the relativistic limit,
while our input parameters $(\gamma_{\rm wall}, \alpha) = (10, 10)$ are not as relativistic.

\item
The last ingredient of our effective theory (the energy and momentum domination by the shock front)
is not a strict condition but rather a qualitative one.

\end{itemize}
Regarding the first point, we plot Fig.~\ref{fig:RH} to check the Rankine-Hugoniot relations in our simulation.
In the left panel, the blue and red points show $\gamma_s^2$ and $2\gamma_{\rm peak}^2$, respectively,
which should coincide with each other in the relativistic limit as in the first relation in Eq.~(\ref{eq:RH}).
The yellow points are the Rankine-Hugoniot relations without the relativistic limit 
(Eq.~(\ref{eq:RH1Beyond}) in Appendix~\ref{app:Beyond}).
We see that all the blue, red, and yellow points coincide well with each other.
On the other hand, the right panel of Fig.~\ref{fig:RH} is to check the second relation in Eq.~(\ref{eq:RH}).
The blue and red points are $\rho_{\rm peak} / \rho_0$ and $(16/3) \gamma_{\rm peak}^2$, respectively,
while the yellow points are the corresponding quantities without the relativistic limit 
(Eq.~(\ref{eq:RH2Beyond}) in Appendix~\ref{app:Beyond}).
We see that the blue and yellow points coincide well with each other.
Also, Fig.~\ref{fig:EOM} checks the relation (\ref{eq:C+C-1}).
The blue and red points are  $\ln \rho'$ and $(7/6) \ln {\gamma^2}'$, respectively.
As it is clear from the derivation, this relation is also subject to the first source of error listed above.
However, we see that the difference between the two quantities $\ln \rho'$ and $(7/6) \ln {\gamma^2}'$ is only about $10\%$.

To summarize this section, we found that our effective theory of shock propagation 
describes the evolution of the system rather correctly.
In the next section, we use it to discuss the implications to the GW production.

\section{Implications to gravitational wave production}
\label{sec:Implications}
\setcounter{equation}{0}

\begin{figure}
\begin{center}
\includegraphics[width=0.46\columnwidth]{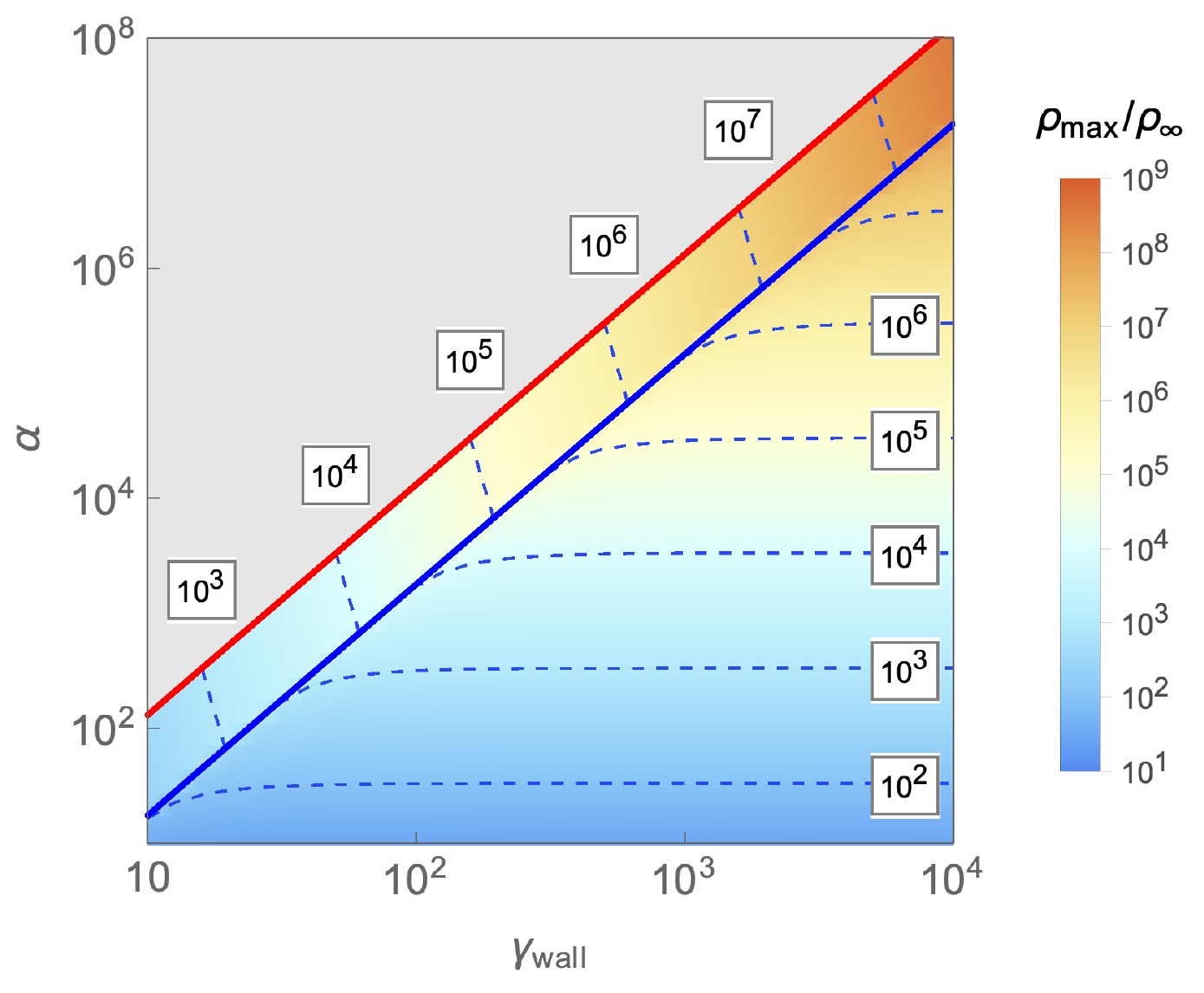}
\hskip 0.4cm
\includegraphics[width=0.46\columnwidth]{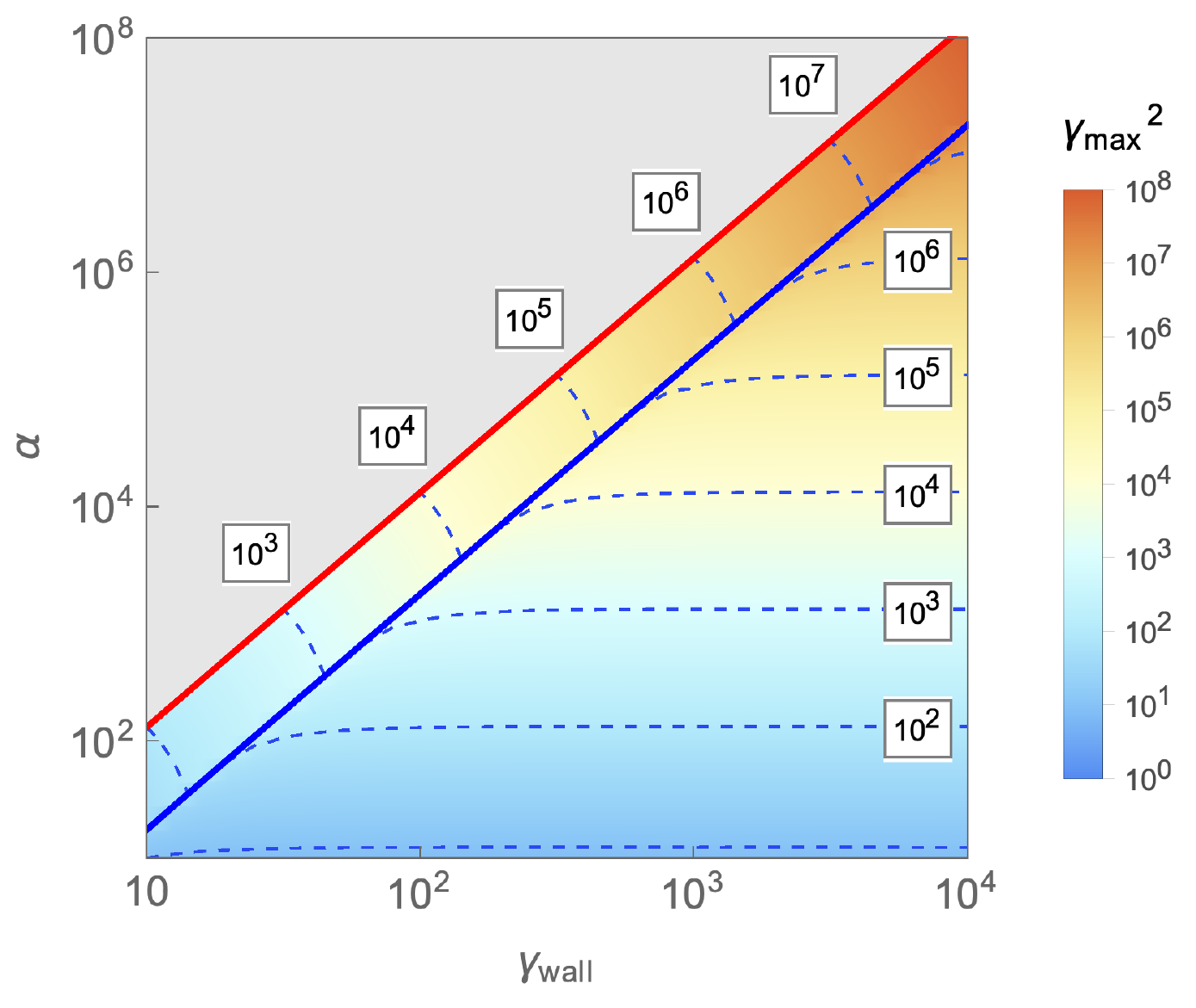}
\hskip 0.1cm
\caption{\small
Maximal fluid energy density $\rho_{\rm max} / \rho_\infty$ (left) 
and $\gamma_{\rm max}^2$ (right) at the wall position before bubble collision.
}
\label{fig:Plane1}
\vskip 2cm
\hskip 0.1cm
\includegraphics[width=0.46\columnwidth]{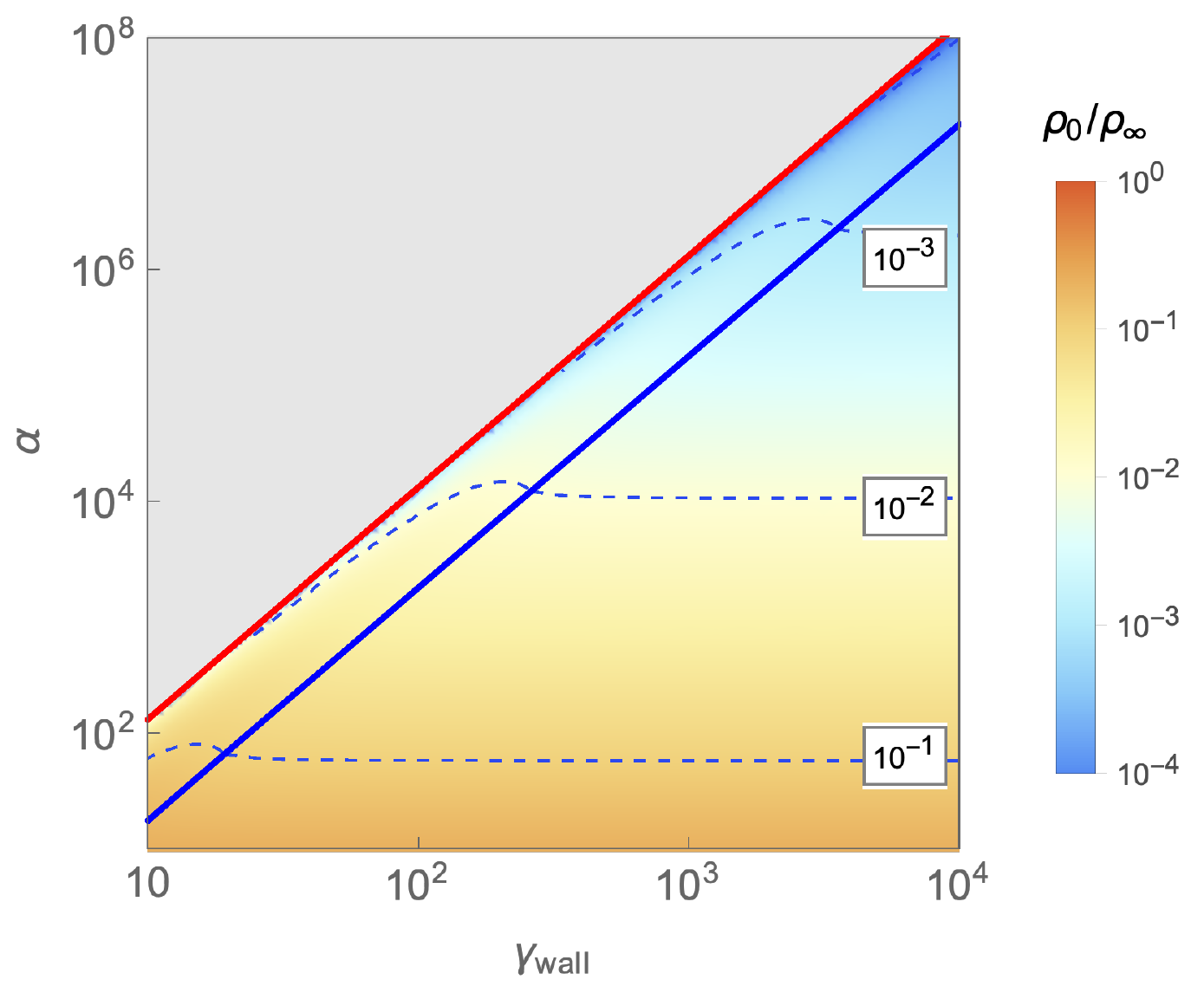}
\includegraphics[width=0.525\columnwidth]{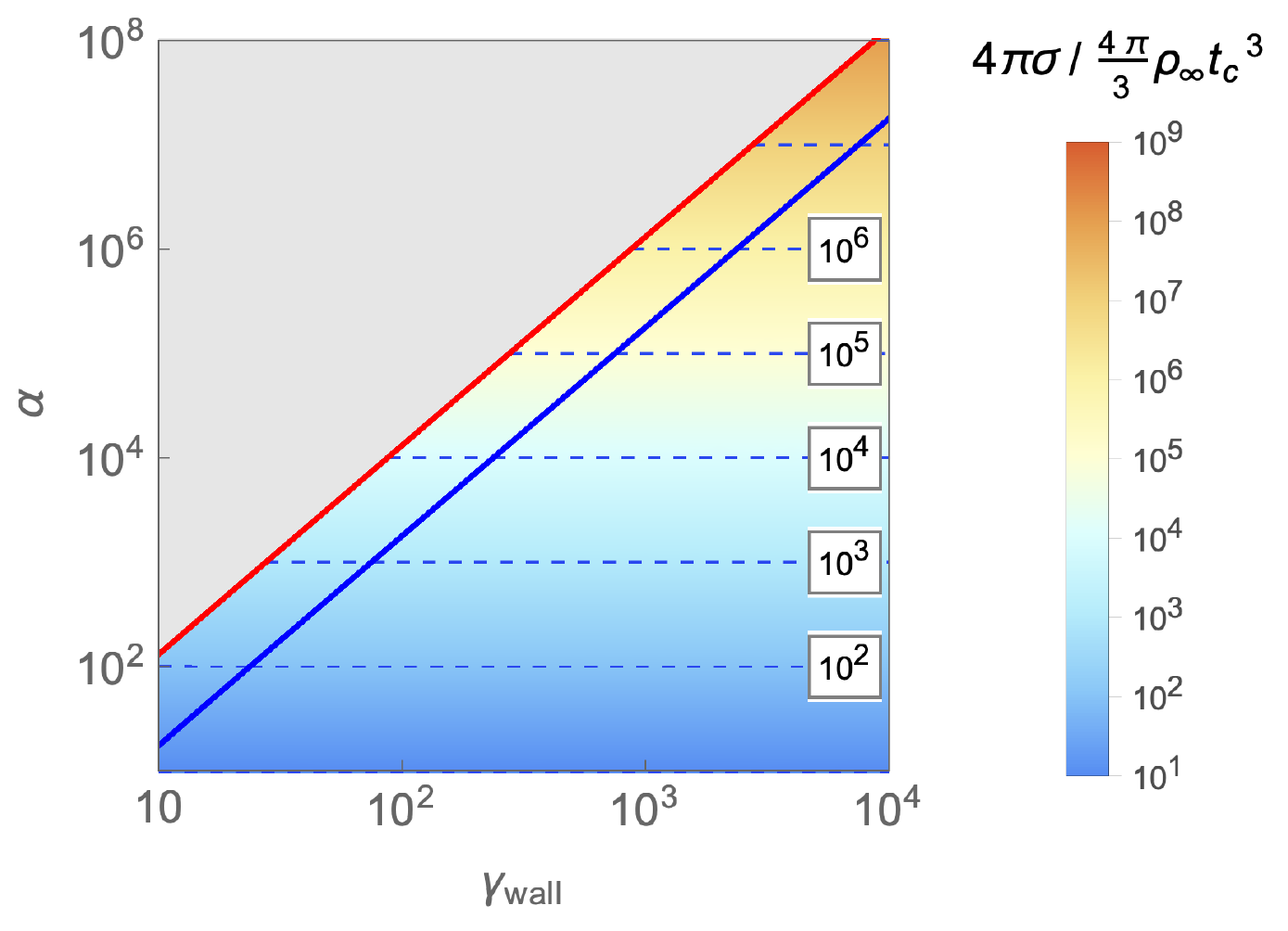}
\caption{\small
Fluid energy density in the broken phase $\rho_0 / \rho_\infty$ during bubble expansion (left) 
and surface momentum density $\sigma$ at the collision time (right).
}
\label{fig:Plane2}
\end{center}
\end{figure}

\begin{figure}
\begin{center}
\includegraphics[width=0.46\columnwidth]{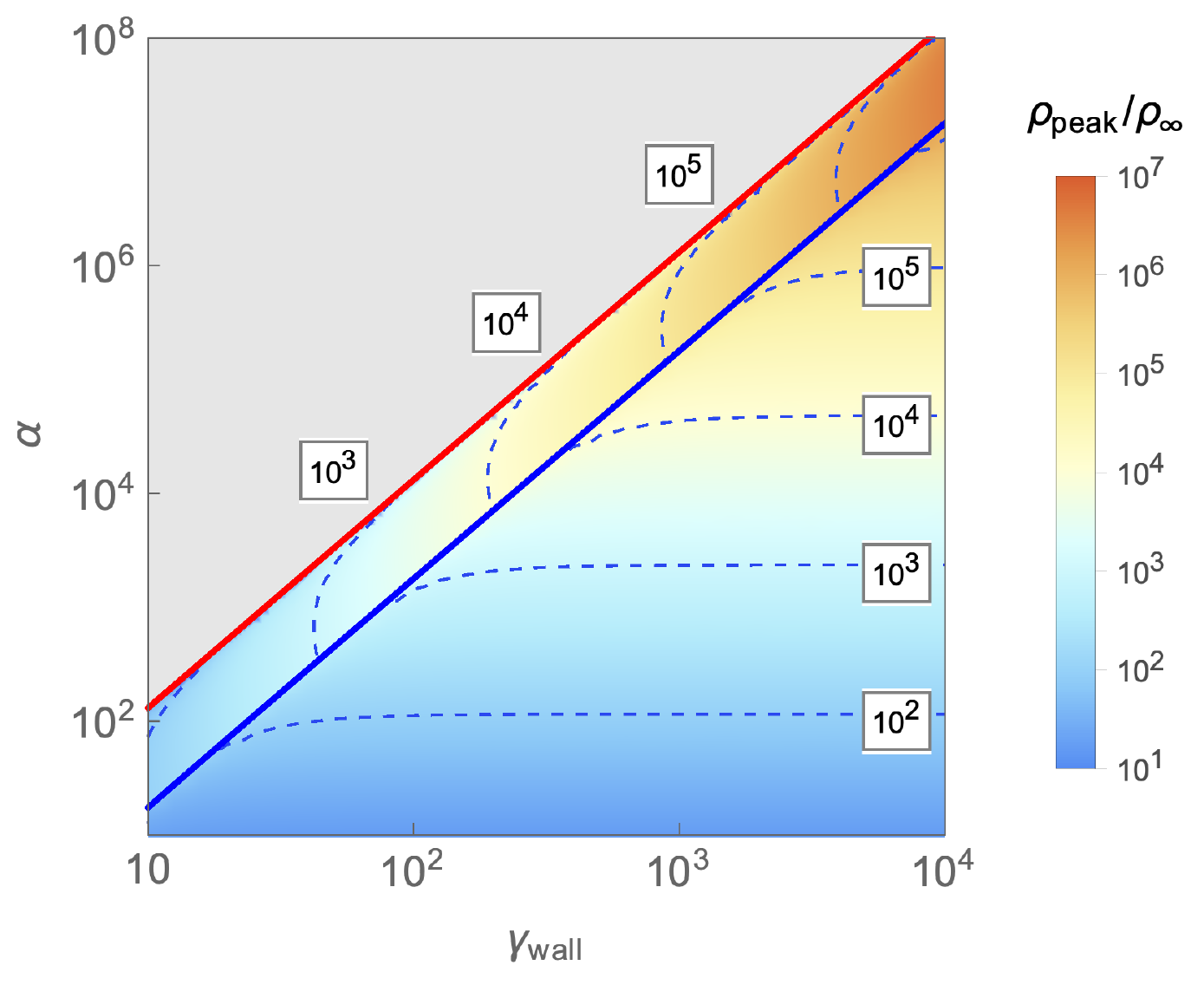}
\hskip 0.4cm
\includegraphics[width=0.46\columnwidth]{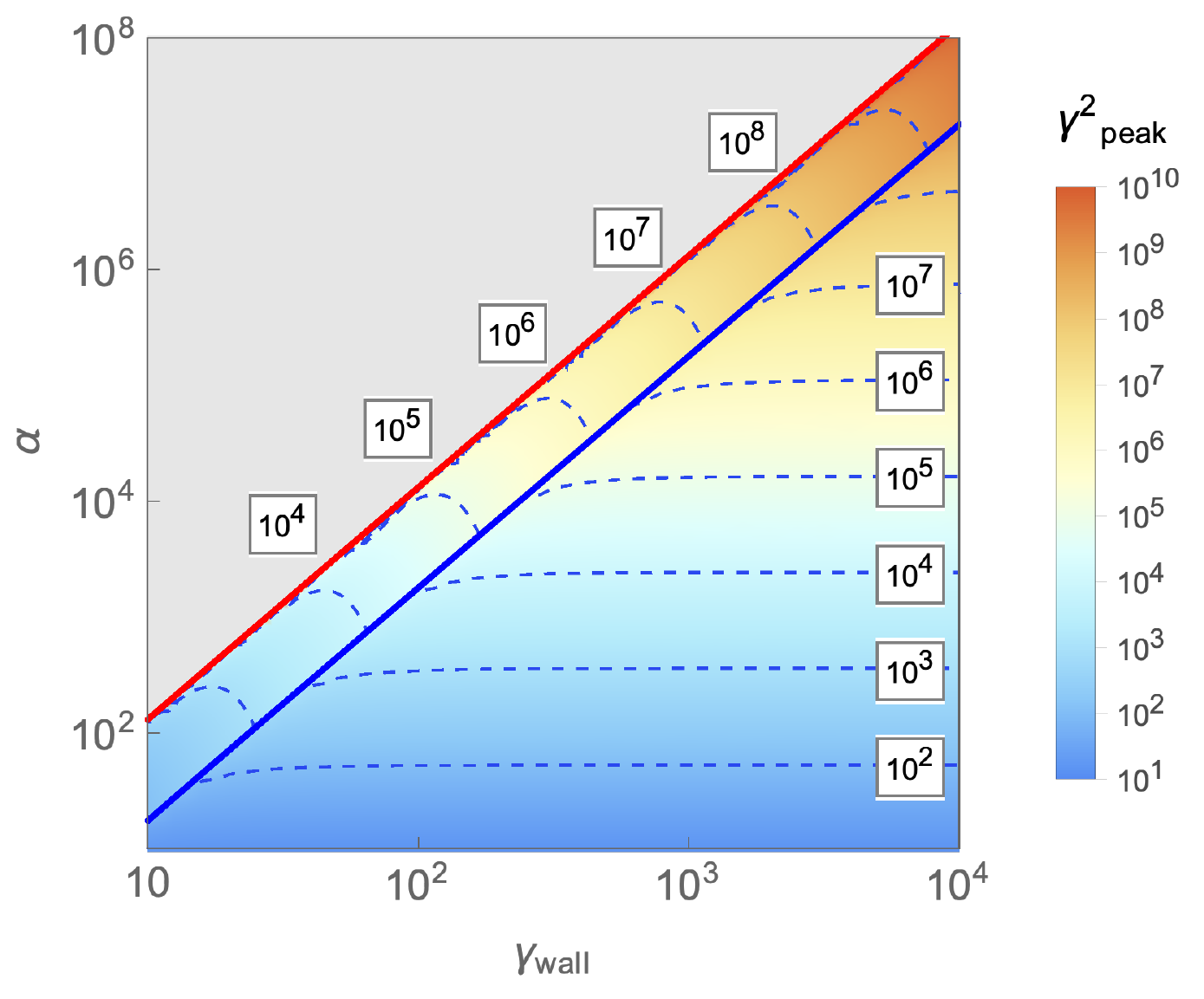}
\hskip 0.1cm
\caption{\small
Initial conditions for the peak fluid energy density $\rho_{\rm peak} / \rho_\infty$ (left) and $\gamma_{\rm peak}^2$ (right)
predicted by the relation (\ref{eq:IC}), together with the Rankine-Hugoniot relations (\ref{eq:RH}).
}
\label{fig:Plane3}
\end{center}
\end{figure}

\begin{figure}
\begin{center}
\includegraphics[width=0.65\columnwidth]{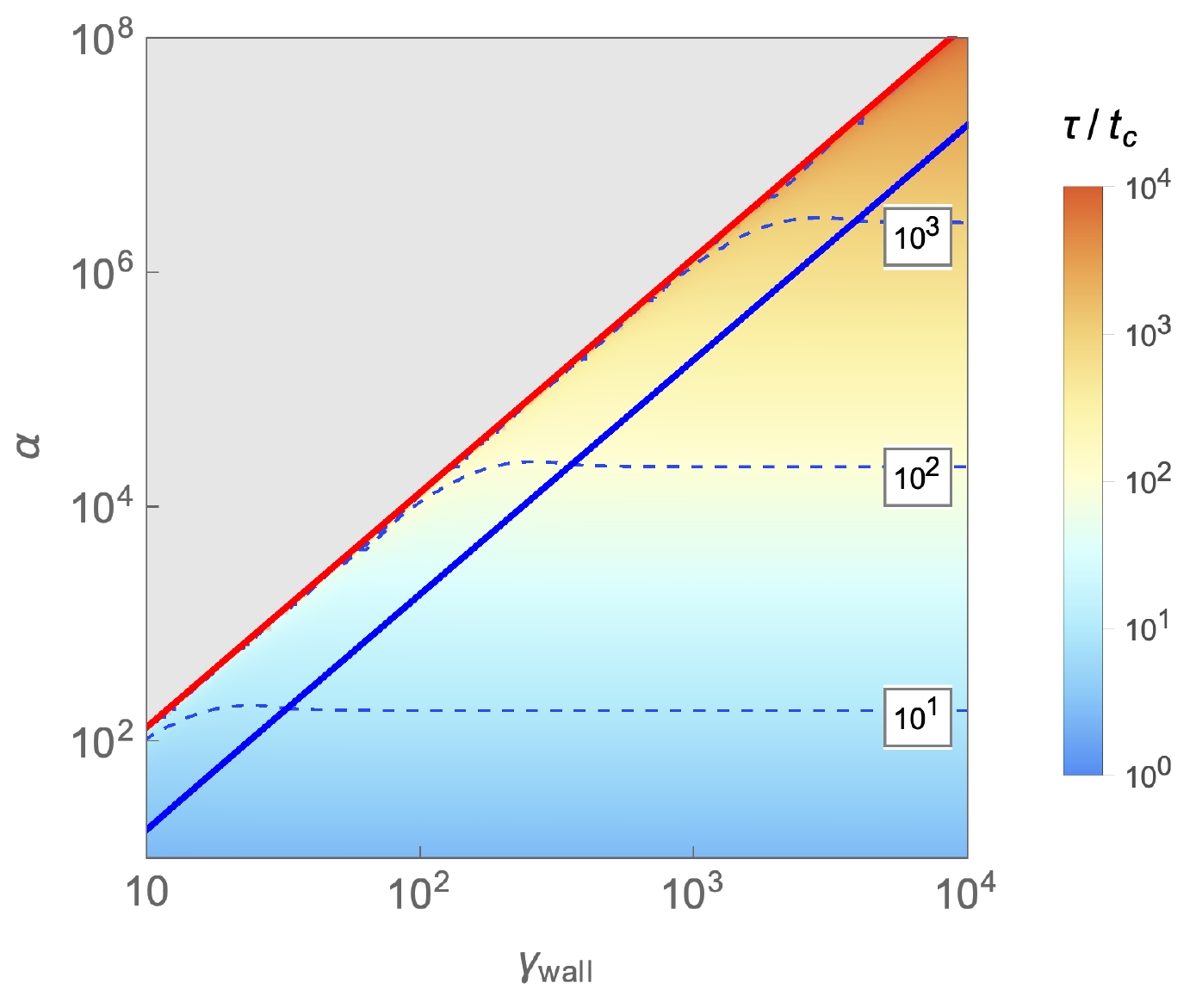}
\caption{\small
Lifetime of the relativistic fluid profiles after bubble collision.
}
\label{fig:tau}
\end{center}
\end{figure}

Now we discuss the implications of our effective theory to the GW production in ultra-supercooled transitions.
In this section, we present our results in the $\gamma_{\rm wall}$-$\alpha$ plane.
In order to translate the $\gamma_{\rm wall}$-$\gamma$ plane (the right panel of Fig.~\ref{fig:Profile}) into this plane, 
we follow the procedure described in Ref.~\cite{Espinosa:2010hh}.
The result is the parameter regions shown in Figs.~\ref{fig:Plane1}--\ref{fig:tau}.
In these figures, the region below the solid blue line is for detonations, 
while the one between the solid red and blue lines is for hybrids.\footnote{
Note that the transition from a detonation to a hybrid is continuous in $\alpha$
in contrast the right panel of Fig.~\ref{fig:Profile} where there is no consistent solution inside the red region.
}
Though the estimate on the initial condition (\ref{eq:IC}) applies only to the detonation profiles
(see footnote~\ref{fn:IC}), we slightly extrapolate our results to the hybrid region,
interpreting $\rho_{\rm max}$ and $\gamma_{\rm max}$ in Eq.~(\ref{eq:IC}) 
as the maximal fluid energy density and $\gamma$ factor in the hybrid case.

Now let us explain each figure. 
Fig.~\ref{fig:Plane1} is the contour plots for $\rho_{\rm max} / \rho_\infty$ (left) and $\gamma_{\rm max}^2$ (right).
The discontinuities across the blue line (corresponding to the boundary between hybrids and detonations) are 
already known~\cite{Espinosa:2010hh} and we successfully reproduce them.
One important observation from this figure is that the contours are parallel to the $\gamma_{\rm wall}$-axis 
in the detonation region.
This means that
\begin{itemize}
\item[]
$\rho_{\rm max}$ and $\gamma_{\rm max}$, which are the input parameters of our effective theory through Eq.~(\ref{eq:IC}),
are $\gamma_{\rm wall}$ independent.
\end{itemize}
This fact may be helpful when estimating the GW production from particle physics models 
since it is often difficult to estimate the wall velocity from microphysics.
Actually, this was already expected from Fig.~\ref{fig:Profile}:
the limit $\xi_{\rm wall} \to 1$ does not cause any divergence in the differential equation (\ref{eq:v_xi}),
and therefore the dynamics does not depend on $\xi_{\rm wall}$ as long as it is close to unity.\footnote{
The authors thank T. Konstandin for pointing this out.
}

Fig.~\ref{fig:Plane2} is the contour plots for $\rho_0 / \rho_\infty$ (left) and $4 \pi \sigma / \frac{4\pi}{3} \rho_\infty t_c^3$ (right).
From the left panel, we see that, for a fixed $\gamma_{\rm wall}$, 
the plasma energy density in the broken phase gets smaller as $\alpha$ gets larger.
This can be understood from the behavior of friction:
the larger the $\alpha$, the larger the pressure on the wall, 
and therefore the friction on the wall increases accordingly in order to keep $\gamma_{\rm wall}$ constant.
The increased friction results in a stronger drag of the fluid behind the wall, thus the energy density inside the bubble gets smaller.
On the other hand, the quantity shown in the right panel is almost the same as $\alpha$.
This is a cross-check of our calculation:
$4 \pi \sigma$ is the total energy in the whole bubble at $t = t_c$
(note that we are interested in the relativistic limit
and the difference between energy and momentum does not matter),
while $\frac{4\pi}{3} \rho_\infty t_c^3$ is the plasma energy density in the symmetric phase 
multiplied by the volume of the bubble at the time of collision.
Therefore, this quantity should coincide with $\alpha$, and it does indeed.

Fig.~\ref{fig:Plane3} is the contour plots for $\rho_{\rm peak}/\rho_\infty$ (left) and $\gamma_{\rm peak}^2$ (right)
calculated from $\rho_{\rm max}$ and $\gamma_{\rm max}^2$ using Eq.~(\ref{eq:IC}) and Eq.~(\ref{eq:RH}).
In these plots, we extrapolated our results to the hybrid region by interpreting 
$\rho_{\rm max}$ and $\gamma_{\rm max}^2$ in Eq.~(\ref{eq:IC}) 
as the maximal energy density and $\gamma$ factor squared of the fluid at the wall position.
Note that, as explained in the beginning of this section, this procedure is not justified in the strict sense 
because the point $W$ in the left panel of Fig.~\ref{fig:Characteristic} does not coincide with the starting point of the shock wave.
Therefore, the contours in the hybrid region should be taken as reference values.

Finally, Fig.~\ref{fig:tau} is a contour plot for the lifetime of shock waves (\ref{eq:tau}).
This is the main result of this paper:
the propagating fluid in the broken phase remains relativistic for quite a long time ($\sim 10^{1,2,3} \times t_c$).
This means that the thickness of the shock wave remains much smaller than the typical bubble size until late times,
and that the onset of the sound wave regime (characterized by the linearity of the fluid equation) is delayed.
However, please note that this conclusion is subject to change
once we include the effects of shock wave collisions.

\section{Discussion and conclusions}
\label{sec:DC}
\setcounter{equation}{0}

In this paper, we studied the fluid dynamics and the resulting GW production in ultra-supercooled first-order phase transitions
characterized by $\alpha \gtrsim 1$.
Though such transitions are observationally and theoretically motivated, 
their evolution and GW production mechanisms are quite unclear.
Such system is likely to proceed with a strong detonation profile~\cite{Bodeker:2017cim}
in which energy gets highly concentrated behind the wall (see Fig.~\ref{fig:Fluid}).
It is extremely hard to tackle this system with numerical simulations
because of the huge hierarchy between the energy concentration scale and the simulation volume,
and also of the strong shock waves forming around the propagating fluid.
Especially, it is unclear when the system develops into the sound wave or turbulence regimes,
the former of which is characterized by the linearity of the fluid equation and is widely believed to serve as a 
long-lasting source of gravitational waves~\cite{Hindmarsh:2013xza,Hindmarsh:2015qta,Hindmarsh:2017gnf,Hindmarsh:2016lnk}.
It is of great importance to clarify this because the GW spectrum of overlapping sound shells can be quite different from that of thin and long-lived bubbles~\cite{Hindmarsh:2016lnk,Jinno:2017fby}.

In order to study the system, we divided the problem into two parts:
the fluid propagation and collision during bubble expansion (Sec.~\ref{sec:Strategy}).
The aim of this paper was to tackle the former.
For this purpose, we developed an effective theory of shock wave propagation (Sec.~\ref{sec:Effective}) valid in the relativistic limit described by a few degrees of freedom (Sec.~\ref{subsubsec:Variables}).
The theory has two constraint equations (the Rankine-Hugoniot relations) 
and two time evolution equations (resulting from the two characteristics),
and becomes closed by assuming the energy or momentum dominance by the shock front
(Sec.~\ref{subsubsec:Equations}).
The effective theory rendered the system analytically solvable (Sec.~\ref{subsubsec:Solution}),
and we checked the validity of the theory by comparing it with numerical simulations in the regime of moderate relativisticity
(Sec.~\ref{subsec:Comparison}).

Equipped with the theory, we finally discussed the implications to the GW production 
in ultra-supercooled transitions in Sec.~\ref{sec:Implications}.
We first found that the initial conditions of the effective theory depend on $\alpha$, 
but not on $\gamma_{\rm wall}$.
This fact can be helpful since it is often difficult to calculate the wall velocity starting from particle physics models.
We also found that the lifetime of the fluid relativisticity can be orders of magnitude larger than the typical bubble size
(Fig.~\ref{fig:tau}).
However, in estimating this lifetime, we took only the propagation effects into account. To further understand the GW production in large $\alpha$ transitions, we need to incorporate the effects of fluid collisions.
This will be addressed in future work.

\section*{Acknowledgments}

The authors are grateful to Thomas Konstandin and G\'eraldine Servant for helpful comments on the manuscript, and particularly grateful to Sangjun Lee for helpful discussions at the initial stage of this project.
The work of RJ was supported by Grants-in-Aid for JSPS Overseas Research Fellow (No. 201960698).
The work of RJ was supported by IBS under the project code, IBS-R018-D1.
The work of RJ is supported by the Deutsche Forschungsgemeinschaft under Germany's Excellence Strategy -- EXC 2121 ``Quantum Universe" -- 390833306.
The work of MT was supported by JSPS Research Fellowships for Young Scientists.

\appendix

\section{Effective theory without the relativistic limit}
\label{app:Beyond}
\setcounter{equation}{0}

In this appendix, we summarize the Rankine-Hugoniot conditions 
and time evolution equations without taking the relativistic limit.
For the constraint (\ref{eq:sigma}), we do not show the corresponding equation 
since Eq.~(\ref{eq:sigma}) is already an approximate relation.

\subsection*{Rankine-Hugoniot conditions}

The Rankine-Hugoniot conditions are $d$-independent and given by
\begin{align}
\gamma_s^2
&= 
\gamma_{\rm peak}^2
\left(
\displaystyle 1 + \frac{1}{2\gamma_{\rm peak}^2}
+
\sqrt{1 - \frac{1}{\gamma_{\rm peak}^2}}
\sqrt{1 - \frac{1}{4\gamma_{\rm peak}^2}}
\right),
\label{eq:RH1Beyond}
\\
\frac{\rho_{\rm peak}}{\rho_0}
&= 
1 
+ 
\frac{8}{3}
\gamma_{\rm peak}^2
\sqrt{1 - \frac{1}{\gamma_{\rm peak}^2}}
\left(
\sqrt{1 - \frac{1}{\gamma_{\rm peak}^2}}
+
\sqrt{1 - \frac{1}{4\gamma_{\rm peak}^2}}
\right).
\label{eq:RH2Beyond}
\end{align}

\subsection*{Time evolution equations}

The time evolution equations are derived in almost the same way as Sec.~\ref{subsubsec:Equations}
except that we do not take the relativistic limit.
For $C_+$ and $C_-$, we have
\begin{align}
&
\partial_t
\left[
\frac{1}{2} \ln \left( \rho_{\rm peak}^{\sqrt{3}/2} \frac{1 + v_{\rm peak}}{1 - v_{\rm peak}} \right)
\right]
\nonumber \\
&~~~~~~
=
- \left( 
v_{C_+}|_{\rm peak} - v_s 
\right)
~
\partial_r
\left.
\left[
\frac{1}{2} \ln \left( \rho^{\sqrt{3}/2} \frac{1 + v}{1 - v} \right)
\right]
\right|_{\rm peak}
-
\frac{v_{\rm peak} c_s}{1 + v_{\rm peak} c_s}
\frac{d - 1}{r_s}
\nonumber \\
&~~~~~~
=
- \left( 
v_{C_+}|_{\rm peak} - v_s 
\right)
~
\partial_r
\left.
\left[
\frac{1}{2} \ln \left( \rho^{\sqrt{3}/2} \frac{1 + v}{1 - v} \right)
\right]
\right|_{\rm peak}
-
\frac{\sqrt{3}}{2}
\left(
v_{C_+}|_{\rm peak} - \frac{1}{\sqrt{3}}
\right)
\frac{d - 1}{r_s},
\end{align}
\begin{align}
&
\partial_t
\left[
\frac{1}{2} \ln \left( \rho_{\rm peak}^{- \sqrt{3}/2} \frac{1 + v_{\rm peak}}{1 - v_{\rm peak}} \right)
\right]
\nonumber \\
&~~~~~~
=
\left( 
v_s - v_{C_-}|_{\rm peak}
\right)
~
\partial_r
\left.
\left[
\frac{1}{2} \ln \left( \rho^{- \sqrt{3}/2} \frac{1 + v}{1 - v} \right)
\right]
\right|_{\rm peak}
+
\frac{v_{\rm peak} c_s}{1 - v_{\rm peak} c_s}
\frac{d - 1}{r_s}
\nonumber \\
&~~~~~~
=
\left( 
v_s - v_{C_-}|_{\rm peak}
\right)
~
\partial_r
\left.
\left[
\frac{1}{2} \ln \left( \rho^{- \sqrt{3}/2} \frac{1 + v}{1 - v} \right)
\right]
\right|_{\rm peak}
+
\frac{\sqrt{3}}{2}
\left(
v_{C_-}|_{\rm peak} + \frac{1}{\sqrt{3}}
\right)
\frac{d - 1}{r_s}.
\end{align}
These give the time evolution equations without the relativistic limit as
\begin{align}
&
\frac{\sqrt{3}}{2} \partial_t \ln \rho_{\rm peak}
+
\frac{\partial_t \ln \gamma_{\rm peak}^2}{\sqrt{1 - 1 / \gamma_{\rm peak}^2}}
\nonumber \\
&~~~~~~
=
- \left( 
v_{C_+}|_{\rm peak} - v_s 
\right)
\left[
\frac{\sqrt{3}}{2} \ln \rho'
+
\frac{\ln {\gamma^2}'}{\sqrt{1 - 1 / \gamma_{\rm peak}^2}}
\right]
-
\sqrt{3}
\left(
v_{C_+}|_{\rm peak} - \frac{1}{\sqrt{3}}
\right)
\frac{d - 1}{r_s},
\label{eq:C+nonrel}
\\
-
&
\frac{\sqrt{3}}{2} \partial_t \ln \rho_{\rm peak}
+
\frac{\partial_t \ln \gamma_{\rm peak}^2}{\sqrt{1 - 1 / \gamma_{\rm peak}^2}}
\nonumber \\
&~~~~~~
=
\left( 
v_s - v_{C_-}|_{\rm peak}
\right)
\left[
- \frac{\sqrt{3}}{2} \ln \rho'
+
\frac{\ln {\gamma^2}'}{\sqrt{1 - 1 / \gamma_{\rm peak}^2}}
\right]
+
\sqrt{3}
\left(
v_{C_-}|_{\rm peak} + \frac{1}{\sqrt{3}}
\right)
\frac{d - 1}{r_s},
\label{eq:C-nonrel}
\end{align}
where
\begin{align}
v_{C_\pm}|_{\rm peak}
&=
\frac{v_{\rm peak} \pm c_s}{1 \pm v_{\rm peak} c_s} \nonumber \\[1ex]
&=
\left(
\sqrt{1 - \frac{1}{\gamma_{\rm peak}^2}} \pm \frac{1}{\sqrt{3}}
\right)
\Big/
\left(
1 \pm \frac{1}{\sqrt{3}} \sqrt{1 - \frac{1}{\gamma_{\rm peak}^2}}
\right),
\\[1.5ex]
v_s
&=
\sqrt{
1 - \frac{4}{9} \left(
1 + \frac{1}{2\gamma_{\rm peak}^2} 
- \sqrt{1 - \frac{1}{\gamma_{\rm peak}^2}} \sqrt{1 - \frac{1}{4\gamma_{\rm peak}^2}}
\right)
}.
\end{align}
Note that $r_s$ in Eqs.~(\ref{eq:C+nonrel}) and (\ref{eq:C-nonrel}) is related to $v_s$ through $v_s = dr_s/dt$
and the initial condition at the bubble nucleation time given by $r_s (t = 0) = 0$.

\section{Details on numerical simulation}
\label{app:Detail}
\setcounter{equation}{0}

\begin{figure}
\begin{center}
\includegraphics[width=0.4\columnwidth]{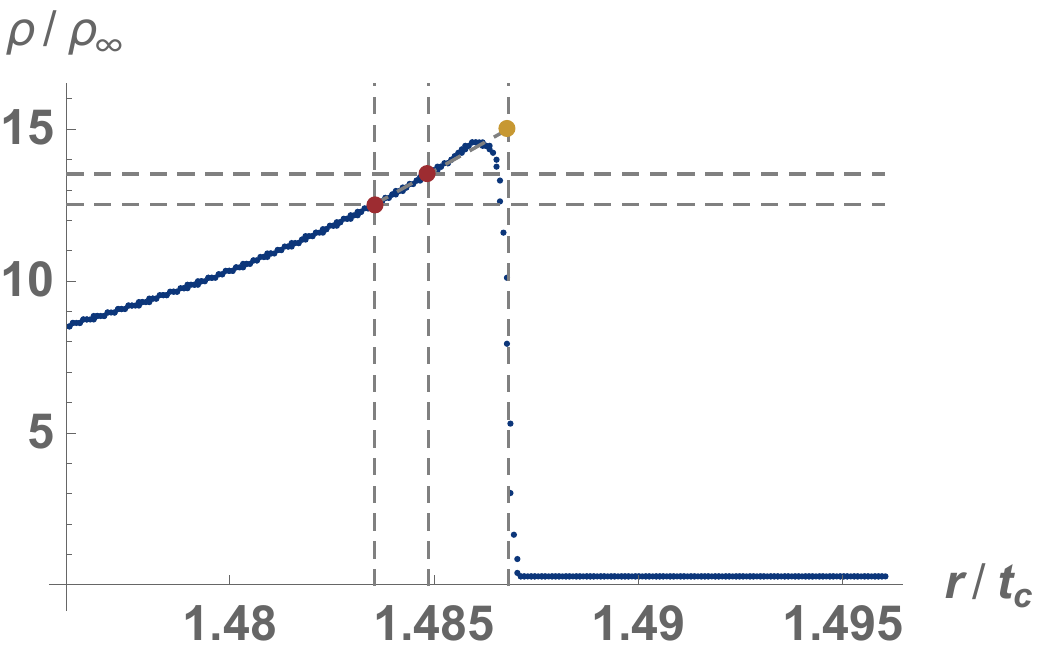}
\includegraphics[width=0.4\columnwidth]{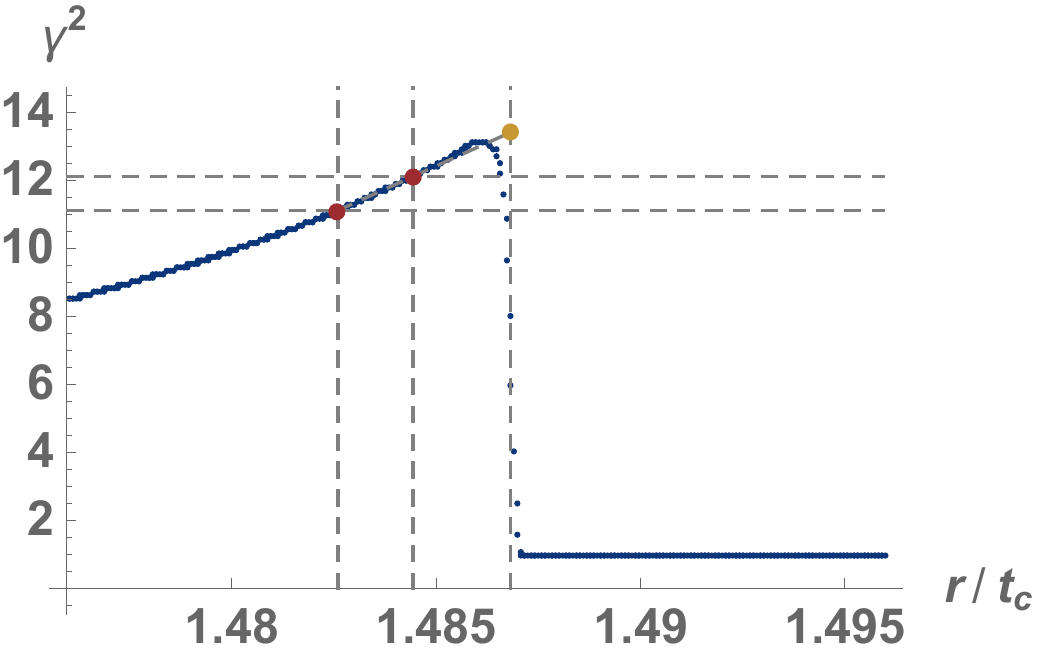}
\caption{\small
The estimation procedure for the shock position and the peak values.
These are snapshots at $t = 1.5 t_c$.
}
\label{fig:Detail}
\end{center}
\end{figure}

In this appendix, we explain how we read off the shock position and peak values from numerical data.
Fig.~\ref{fig:Detail} is a snapshot of the numerical simulation explained in Sec.~\ref{subsec:Comparison} at $t = 1.5 t_c$.
As seen from the figure, both quantities are smeared around the peak 
due to the numerical scheme we adopted and the finiteness of the number of grids.
We first identified the position where the absolute value of the spatial derivative of $\rho$ or $\gamma^2$ 
becomes the largest as the true shock position (yellow points).
The shock position thus estimated is nearly identical for $\rho$ or $\gamma^2$.
We next identify the two points where the value of $\rho / \rho_\infty$ or $\gamma^2$ is 
smaller than the peak value by $1$ or $2$ (red points).
Then we finally extrapolate these points to the estimated shock position
to estimate the true peak values of $\rho$ and $\gamma^2$
as well as the derivatives $\ln \rho'$ and $\ln {\gamma^2}'$.

\clearpage

\small
\bibliography{ref}

\end{document}